\documentclass[seceq]{ptptex}

\usepackage{graphicx}

\usepackage{wrapft}

\title{
  New Generation of the Monte Carlo Shell Model 
  for the K Computer Era
}
\author{%       %Use \scshape  for the family name.
  Noritaka  \textsc{Shimizu}$^{1,}$
  \footnote{E-mail: shimizu@cns.s.u-tokyo.ac.jp} , 
  Takashi  \textsc{Abe}$^{1}$, 
  Yusuke  \textsc{Tsunoda}$^{2}$, 
  Yutaka  \textsc{Utsuno}$^{3}$, 
  Tooru  \textsc{Yoshida}$^{1}$,
  Takahiro  \textsc{Mizusaki}$^{4}$,
  Michio  \textsc{Honma}$^{5}$,
  and Takaharu \textsc{Otsuka}$^{1,2,6}$
}

\inst{
  $^1$Center for Nuclear Study, University of Tokyo, Tokyo 113-0033, Japan\\
  $^2$Department of Physics, University of Tokyo, Tokyo 113-0033, Japan\\
  $^3$Advanced Science Research Center, Japan Atomic Energy Agency,
  Tokai, Ibaraki 319-1195, Japan \\
  $^4$Institute of Natural Sciences, Senshu University, Tokyo,
  101-8425, Japan \\
  $^5$Center for Mathematical Sciences, University of Aizu, Ikki-machi,
  Aizu-Wakamatsu, Fukushima 965-8580, Japan \\
  $^6$ National Superconducting Cyclotron Laboratory,
  Michigan State University, East Lansing, Michigan, USA
}

\abst{
  We present a newly enhanced version of the Monte Carlo Shell Model method 
  by incorporating the conjugate gradient method and 
  energy-variance extrapolation. 
  This new method enables us to perform large-scale shell-model calculations 
  that the direct diagonalization method cannot reach.
  This new generation framework of the MCSM provides us with a powerful
  tool to perform most-advanced large-scale shell-model calculations 
  on current massively parallel computers such as the K computer. 
  We discuss the validity of this method in {\it ab initio} calculations 
  of light nuclei, and propose a new method to describe the intrinsic 
  wave function in terms of the shell-model picture.
  We also apply this new MCSM to the study of neutron-rich Cr and Ni isotopes 
  using the conventional shell-model calculations 
  with an inert $^{40}$Ca core and discuss 
  how the magicity of $N=28, 40, 50$ remains or is broken.
}

\begin{document}
\maketitle

%%%% Sect.1 Introduction, Utsuno
%%% % \input{sect1_utsuno.tex}
% \input{sect1_2.tex}
% Introduction 
\section{Introduction}
\label{introduction}

The understanding of the structure of all nuclei from the first 
principle, called usually the {\it ab initio} nuclear calculation, 
is one of the ultimate goals in nuclear theory.  
For this purpose, one usually starts with the nucleon degree of freedom, 
i.e., protons and neutrons as the building block of a nucleus, 
assuming the free (or bare) nucleon-nucleon force to be the interaction 
between nucleons.  
Recent {\it ab initio} calculations often include 
not only the two-nucleon force but also the three-nucleon 
force which is specific to the interaction between composite 
particles like nucleons. 
Several {\it ab initio} methods have been proposed, and 
their accuracy has been investigated in great detail, for instance, 
in terms of the binding energy of the four-nucleon system 
\cite{kamada01}.  It remains, however, rather difficult or infeasible 
to go beyond systems with $A\gtrsim 12$, 
where $A$ is the number of nucleons. 
This is largely due to strong nucleon-nucleon correlations 
that requires a large number of single-particle 
states to be included for the description of many-body states
in ground states or low excited states.
Although this problem can be resolved to a certain extent by using 
effective interactions based on various renormalization techniques, 
the number of {\it many-body} states to be included remains  
prohibitively large in most cases, and increases exponentially with 
the nucleon number. 
In this paper, we present a new version of the Monte Carlo shell model
(MCSM), demonstrating how it can contribute to
{\it ab initio} nuclear calculation as well as to 
conventional but quite huge shell model calculations. 

The nuclear shell model has been known to be successful in 
describing the structure of atomic nuclei based on nuclear forces. 
It dates back to 1949 when Mayer and Jensen discovered 
the magic numbers as a consequence of   
a mean potential including the spin-orbit coupling \cite{mj1949}. 
While the original concept of the shell model of Mayer and Jensen 
is basically an independent particle picture, the concept 
has been modified and extended significantly over decades since then. 
The modern shell model uses a sufficiently large number 
of many-body basis states which are superposed utilizing properly constructed 
effective interactions so as to provide us with accurate many-body
eigenstates.  The many-body basis state is
a Slater determinant usually with single-particle states taken 
from a harmonic oscillator potential.   

In conventional shell-model calculations, an inert core is 
assumed: all single-particle states below a given magic number  
are completely occupied, forming a closed shell. 
Single-particle states between this magic number and the next
magic number constitute a {\it valence shell}, and nucleons 
in this shell are called valence nucleons.
In the conventional shell model, only valence nucleons are 
activated.  The single-particle states of activated nucleons 
in the shell-model calculation are called the {\it model space}.
The model space is a concept for calculation, and is the same as  
some valence space in many cases.  But, in other cases, the
model space can be taken wider or smaller than the relevant 
valence shell depending on some interest or limitation.  
We thus distinguish model space from valence shell hereafter. 
Note that the model space is a more computational concept.

An effective interaction is obtained for valence nucleons, and
is defined for each model space.  
Effects from the inert core, those from virtual 
excitations from the inert core and those from virtual 
excitations to states above the model space are assumed to 
all be incorporated into this effective interaction by renormalizing  
it appropriately.  This can be a very important issue, but the
conventional shell model assumes that there is such an interaction, 
while its determination can be phenomenological.

It has been shown that many nuclear properties 
with $8\le N(Z) \le 20$ are excellently described 
with the conventional shell model calculation 
in the $1s$-$0d$ (model) space 
(often called the $sd$ shell) \cite{usd, usdab}, 
and those with $20\le N(Z)  \lesssim 32$ 
are also quite well described 
in the $1p$-$0f$ (model) space (often called the $pf$ shell) 
\cite{kb3g, gxpf1}. 
Note that the effective interactions used in those calculations
are hybrid products of microscopic derivation and empirical
adjustments. 

While the conventional shell model assumes a relatively limited model 
space as exemplified above, the same computational procedure 
is applicable to the {\it ab initio} calculation 
when no inert core is assumed and a large number of single-particle 
states are taken 
so that the calculation becomes close to calculations in the whole 
Hilbert space. 
Here, we refer to this {\it ab initio} method within the shell model
as the {\it ab initio} shell model, one of which is the no-core  
shell model (NCSM) \cite{NCSM}, a famous model with great success. 

Whether the {\it ab initio} shell model 
or the conventional shell model (with a core) is used, 
all one has to do in computation is 
to diagonalize a Hamiltonian matrix spanned by all the possible 
many-body states in a given model space ({\it i.e.}, 
single-particle space of activated nucleons).
The number of relevant many-body basis states determines 
the dimension of this matrix.  This dimension is often called 
the shell-model dimension, and causes a serious 
computational issue as we shall see later.  
The number of the many-body states is 
$_{N_p}C_{n_p} \times\, _{N_n}C_{n_n}$ 
without symmetry consideration, where 
$N_p$ ($N_n$) is the number of proton (neutron) single-particle states taken 
and $n_p$ ($n_n$) is the number of protons (neutrons) 
activated.  It 
roughly increases exponentially with $N_p$ (or $N_n$) 
and also with $n_p$ (or $n_n$). 
Hence, without even considering the {\it ab initio} shell model
with a huge number of $N_p$ and $N_n$, 
the conventional shell model for heavier nuclei already suffers 
from the huge dimensionality 
of the Hamiltonian matrix necessary to tackle the whole nuclear chart 
because the number of valence orbits ($N_p$ or $N_n$) increases
for heavier nuclei. For instance, the dimension needed for $A\sim 80$ 
$N=Z$ nuclei is estimated to be  $\sim 10^{27}$ 
without any symmetry consideration \cite{scidac}
when the $1p$-$0f$-$2s$-$1d$-$0g$ orbits are assumed to be the valence 
shell. 
Although this number can be reduced 
by 1-2 orders of magnitude by choosing only the states of the same $J_z$ 
as taken usually 
(i.e., $M$-scheme calculation), 
it is still far from the current computational limit of 
$10^{10-11}$ in the $M$-scheme. % \cite{dimension}. 

In order to go beyond the computational limit of the shell model, 
a new method 
for performing the shell-model calculation named 
the Monte Carlo shell model (MCSM) 
has been developed 
since 1996 \cite{mcsm_1996}, 
guided by the Quantum Monte Carlo Diagonalization method \cite{mcsm_1995}. 
The MCSM utilizes the idea of the auxiliary field Monte Carlo method 
that is taken in the Shell Model Monte Carlo method \cite{smmc}, 
but this method and the MCSM are completely different. 
The MCSM aims to represent many-body states with 
a small number of highly selected many-body basis states. 
In this sense, the MCSM is regarded as an ``importance truncation'' of 
the entire many-body space \cite{ppnp_mcsm}. 
The basis state should be represented in a compact form (i.e., 
a wave function with a small number of parameters), 
it should be able to approximate the nuclear many-body state efficiently, 
and its matrix elements should be calculated easily. 
To meet this demand, we usually use deformed 
Slater determinants, and parity and total angular momentum projection
are operated on them if needed. 
Otherwise, a pair condensed state \cite{mcsm_pair} 
or a quasi-particle vacuum state, which is used 
in the Hartree-Fock-Bogoliubov calculation 
and also taken as the basis state of the VAMPIR 
(abbr. for Variational After Mean field Projection In Realistic model spaces)
method \cite{vampier_2004}, 
is a good candidate when required. 
Note that the VAMPIR method has been developed with more emphasis 
on calculating the energy spectra rather than 
improving the energy of a specified state. 
The MCSM basis states are added one by one, and each basis state to be added 
is selected among many candidate Slater determinants generated 
stochastically so that the energy of the state under consideration 
can be as low as possible. 
This step is repeated until the energy converges sufficiently. 
Thus, this original MCSM is characterized as ``stochastic'' 
in terms of the method of basis variation, and as ``sequential'' 
in terms of the procedure of basis variation. 

From the computational point of view, the MCSM has an advantage 
over the conventional diagonalization method 
in tolerance to the increase of the model space and the particles. 
When it is assumed that the number of basis states needed 
in a MCSM wave function is fixed, the total computational cost is 
scaled by the cost of each Hamiltonian matrix element. 
This is roughly proportional to $(N_{sp})^{\alpha}$ with $\alpha\sim 3$-$4$ 
in the case of $N_{sp}=N_p=N_n$, being much milder than the exponential
increase. 
This advantage had enabled us to perform the full $pf$-shell calculation 
in $^{56}$Ni (with $\sim10^9$ $M$-scheme dimension) 
with good accuracy \cite{mcsm_1998} 
several years before the exact diagonalization was carried out
\cite{vampier_2004}.  
See the review paper \cite{ppnp_mcsm} for more details of 
early achievements.

Recently, the computational environment has been changing rapidly. 
The number of available CPU cores is expanding 
to be typically in the range of tens of thousands 
for the world leading supercomputers 
as compared to a few tens to hundreds of CPU cores used in the early 
MCSM calculations. 
The K computer will contain more than 700,000 
cores upon its completion in the autumn of 2012. 
This situation has strongly motivated us to renew the MCSM method  
to be suitable using up-to-date massive parallel computers, 
in addition to wider applications including {\it ab initio} shell model  
calculations. 
%A new MCSM package has been developed since 2009 \cite{riken_rmcsm}. 
%It is renewed not only in terms of coding 
%but also in terms of methodology and numerical algorithm. 
% Thus, it is a package.
Since 2009, we have developed a new MCSM package, which includes 
not only renewed code, but also a novel methodology and numerical algorithm
\cite{riken_rmcsm}.
Among the methodological advancements, the most important is 
the introduction of the energy-variance extrapolation method \cite{mcsm_extrap}. 
In the original MCSM, the energy of a many-body 
state is evaluated directly from the energy expectation value 
of the MCSM wave function,  
which must be higher than the exact solution.  
It was quite hard to know the difference between this value and 
the exact value. 
The energy-variance extrapolation method serves as a powerful 
tool to pin down accurately the exact solution from 
a series of approximated solutions. 
Another methodological change is incorporating a variational aspect 
into the MCSM by varying the basis state, which enhances the
lowering of calculated energy values.
Equipped with the variational-type improvement,  
the MCSM is now characterized as ``deterministic'' in terms of 
the method of basis variation.  
Thus, keeping the original idea, 
the present MCSM, associated with several advancements, can be 
regarded as a new generation.

In this paper, we review the outline of the new-generation MCSM and 
show some of its earliest applications that were not feasible  
with the original MCSM. 
This paper is organized as follows.
In Sect. \ref{sect2_shimizu},  the outline of the new-generation MCSM 
and its feasibility are presented. 
We also discuss the adaptability of the MCSM to massively parallel
supercomputers. 
In Sect. \ref{sect3_tabe}, application to the {\it ab initio} shell
model is demonstrated. Along with numerical success, 
the intrinsic shape and the clustering of light nuclei are also discussed 
with the MCSM wave function. 
In Sect. \ref{sect4_ytsunoda}, application to the neutron-rich chromium 
and nickel isotopes is presented as a case of medium-heavy mass nuclei. 
This region is being intensively studied in radioactive isotope
facilities over the world, and also challenges nuclear models 
because several shapes coexist and are mixed, 
which results in the rapid change of the dominant shape over the isotopes.
In Sect. \ref{sect5_summary}, we summarize this paper and give 
an outlook and future perspective of the MCSM toward the launch of the K computer. 

We note here that many new features may appear in the structure 
of exotic nuclei, because unbalanced numbers of protons and 
neutrons may create situations where unknown or little known 
aspects of nuclear forces become visible and produce large
impacts.  The shell evolution due to the tensor force
\cite{otsuka_tensor_2001, otsuka_tensor_2005, otsuka_tensor_2010}
and the three-body force \cite{otsuka_threebody_2010} are some examples.
Note that the shell evolution explains basic trends of single-particle 
properties and one needs comprehensive calculations to obtain physical 
quantities and look into correlations in depth.
The exploration of such unknown features need theoretical
framework directly linked to nuclear forces, and we expect a 
large contribution from the new generation of MCSM.

%%%% Sect.2 Method, Shimizu
%\input{sect2_shimizu.tex}

\section{Outline of the new-generation Monte Carlo shell model}
\label{sect2_shimizu}

The new-generation MCSM can be 
divided roughly into two stages:
the variational procedure to obtain the approximated wave function 
and the energy-variance extrapolation 
utilizing the obtained wave function. 
We briefly describe these two parts 
in Sects. \ref{sec:var} and \ref{sec:eve}, respectively. 
The shell-model Hamiltonian and the form of the variational wave function 
used in the MCSM is shown in Sect.\ref{sec:wf}.
The additional improvement to make the extrapolation procedure stable
is demonstrated in Sect. \ref{sec:eve_ni56}. 
The numerical aspects of the framework mainly for 
massively parallel computation are discussed in Sect. \ref{sec:computational_aspects}.

\subsection{Shell-model Hamiltonian and variational wave function}
\label{sec:wf}

In conventional nuclear shell-model calculations assuming an inert core, 
we use a general two-body interaction:
\begin{equation}
  H = H^{(1)} + H^{(2)}= \sum_{i} t_{i} c^\dagger_i c_i
  + \sum_{i<j, k<l} v_{ijkl} c^\dagger_i c^\dagger_j c_l c_k ,
\end{equation}
where $c^\dagger_i$ denotes a creation operator of
single particle state $i$.
$H^{(1)}$ is a one-body Hamiltonian with single-particle energies $t_i$, 
and $H^{(2)}$ is a two-body interaction which has parity and rotational symmetry and 
is represented by so-called Two-Body Matrix Elements (TBMEs) \cite{ppnp_brown}.

In the case of {\it ab initio} shell-model calculations, 
the Hamiltonian is taken as
\begin{equation}
  H = T -T_{\rm CM} + V = \sum_{ij} t_{ij} c^\dagger_i c_j
  + \sum_{i<j, k<l} v_{ijkl} c^\dagger_i c^\dagger_j c_l c_k ,
  \label{eq:hamiltonian}
\end{equation}
where $T$ and $T_{\rm CM}$ are the total kinetic energy and 
the kinetic energy of the center-of-mass motion, respectively.
Note that $T_{\rm CM}$ has both one-body and two-body components.
The $V$ represents two-nucleon interaction, e.g., 
the JISP16 interaction \cite{Shirokov07} in Sect. 3. 
In this work, because we do not treat explicitly three-nucleon forces, 
both the Hamiltonians of these two kinds of shell-model calculations 
consist of one-body and two-body interactions.

If necessary, the removal of spurious center-of-mass motion can be done 
by utilizing the prescription of Gloeckner and Lawson \cite{lawson} 
to suppress the contamination of the center-of-mass motion. 
In this prescription the Hamiltonian to be diagonalized is replaced by
\begin{equation}
  \label{eq:lawson}
  H' = H + \beta_{\rm cm} H_{\rm cm} 
\end{equation}
with $H_{\rm cm}$ being defined as 
\begin{equation}
  \label{eq:hcm}
  H_{\rm cm} = \frac{\bf P^2}{2AM} + \frac12 MA\omega^2 {\bf R}^2 
  - \frac32 \hbar \omega, 
\end{equation}
where $\bf R$ and $\bf P$ are the position and momentum of the center of mass.
By taking $\beta_{\rm cm}$ large enough, 
$\langle H_{\rm cm}\rangle $ is suppressed as a small value.

In the framework of the MCSM, 
the approximated wave function is written as a linear combination of 
angular-momentum- and parity-projected Slater determinants,
\begin{equation}
  |\Psi_{N_b} \rangle = \sum_{n=1}^{N_b} \sum_{K=-I}^I f^{(N_b)}_{n,K} P^{I \pi}_{MK}
  | \phi_n \rangle , 
  \label{eq:wf}
\end{equation}
where $N_b$ is the number of the Slater-determinant basis states.
The $P^{I\pi}_{MK}$ operator is the angular-momentum and parity projector defined as
\begin{equation}
  P^{I\pi}_{MK} =  \frac{1 + \pi \Pi}{2}\frac{2I+1}{8\pi^2} 
  \int d\Omega \  {D^{I}_{MK}}^*(\Omega)
  e^{i\alpha J_z} e^{i\beta J_y} e^{i\gamma J_z} ,
  \label{eq:proj}
\end{equation}
where $\Omega \equiv (\alpha,\beta,\gamma)$ are the Euler angles
and $D^I_{MK}(\Omega)$ denotes Wigner's $D$-function. $\Pi$ stands for 
the parity transformation.
Each $|\phi_n\rangle$ is a deformed Slater determinant defined in Eq.(\ref{eq:slater}).
The coefficients $f^{(N_b)}_{n,K}$ are determined 
by the diagonalization of the Hamiltonian matrix in the subspace spanned 
by the projected Slater determinants, $P^{I\pi}_{MK} | \phi_n \rangle $
with $-I \leq K \leq I$ and  $1\leq n\leq N_b$.
This diagonalization also determines the energy, 
$E_{N_b} \equiv \langle \Psi_{N_b} | H |\Psi_{N_b} \rangle$, as a function of $N_b$.
Note that the dimension of the subspace is $(2I+1)N_b$, not $N_b$.
% a product of the number of states, $N_b$, 
% and the degree of freedom of the $z$-component of angular momentum, $2I+1$.
The Slater determinant basis,  $D^{(n)}$, is given by variational calculation
to minimize  $E_{N_b=n}$. 
We increase $N_b$ until $E_{N_b}$ converges enough, 
or the extrapolated energy converges. 
The strategy of this variational calculation 
will be discussed in the next subsection.

\subsection{Variational procedure}
\label{sec:var}

% To describe the efficient computation used in the present study, 
In this section, 
we discuss the efficient process for the determination of $D^{(n)}$
in Eq.(\ref{eq:slater}) to minimize $E_{N_b}$ 
and the history of its developments in this section. 
In the original MCSM calculation, 
the basis states are selected from many candidates generated stochastically
utilizing the auxiliary-field Monte Carlo technique, 
the detail of which was discussed in Ref. \citen{ppnp_mcsm}.
In order to determine $D^{(n)}$ in Eq.(\ref{eq:wf})  efficiently,  
M. Honma {\it et al.} 
introduced the few-dimensional basis approximation \cite{fda}, 
in which they adopted the steepest gradient method. 
It succeeded in estimating the energy of $pf$-shell nuclei with 
a relatively small number of bases, 
however, the direction of the gradient is not necessarily 
the most efficient choice to reach the local minimum and it is 
practically difficult to control the step width in the gradient direction.
Mainly because of this problem, the number of basis states 
was limited to be rather small ($N_b \simeq 30$) compared to that in the MCSM 
($N_b \simeq 100$).
On the other hand, W. Schmid {\it et al. } adopted 
the quasi-Newton method for energy minimization 
in the VAMPIR approximation
to obtain the optimized quasi-particle vacua as basis states  \cite{ppnp_schmid}. 
G. Puddu demonstrated that the quasi-Newton method works also in the 
Slater determinant bases. 
In the quasi-Newton method, 
the step width is automatically determined to minimize the energy 
along the modified gradient direction \cite{hybrid_h2,puddu_eqs}. 
As a result, the step width is relatively large in early steps of 
the quasi-Newton procedure along the sophisticated gradient 
direction so that the additional basis state 
has sufficiently
linearly independent component of the other basis states. % which are already optimized.

In the present work, we adopt 
the conjugate gradient (CG) method \cite{num_recipe, cg_egido}, 
which also includes the linear minimization in the modified gradient direction.
In comparison to the quasi-Newton method, the CG method 
is expected to be advantageous to save memory usage 
and improve computational efficiency for parallel computation 
because a large Hessian matrix is utilized in the quasi-Newton method, 
which is not used  in the CG method.
Note that the energy of the linear combination of 
projected Slater determinants is 
optimized, not that of the unprojected Slater determinants. 
In this sense, this variational procedure is ``variation after projection and 
configuration mixing''.

We proposed the following four steps to optimize 
the parameters of deformed Slater determinants: 
\begin{enumerate}
  \item Monte Carlo sampling utilizing auxiliary field technique (original MCSM), 
  \item Sequential optimization for each basis state (SCG), 
  \item Refinement process of each basis states repeatedly (refinement), 
  \item Full (simultaneous) optimization of all basis states (FCG).
\end{enumerate}

In the first step, 
we perform the original MCSM procedure to obtain the approximated 
wave function, 
which can be used as an initial state of the CG process.
The detail of the original MCSM is not discussed here, but in Ref. \citen{ppnp_mcsm}.
In Refs. \citen{ppnp_mcsm, mcsm_extrap}. we select the best basis states 
from an order of 1000 candidates, which is generated stochastically.
However, the necessary number of candidates can be far suppressed 
if we proceed to the next step.

In the second step, called the SCG method, 
we perform a variational calculation,
with variational parameters being $D^{(i)}$, 
sequentially by minimizing the $E_{i}$ without changing 
the other bases, $D^{(1)}, D^{(2)}, ... D^{(i-1)}$, which were already fixed.
In other words, we perform variational calculations with 
a set of variational parameters, $D^{(i)}$, sequentially.
We increase the number of basis states, $N_b$, until the energy reaches 
sufficient convergence.

In the third step
which is called the refinement process, we take an initial state from the 
result of the SCG calculation. 
In the SCG calculation, 
the $D^{(1)}$ is not the best optimized parameters 
to minimize  $E_{N_b}$, 
since $D^{(1)}$ is determined to minimize $E_1$.
Then, we first fix the number of basis states, $N_m$, 
and restart to minimize the energy $E_{N_m}$
by the CG method to optimize each basis state 
from $D^{(1)}$ to $D^{(N_m)}$, one by one. 
We iterate a few times to perform a routine of 
the CG process for all basis states one by one to get 
a better energy calculation. 

In the fourth step called the FCG, we fix  $N_m$ at the beginning,
and determine all coefficients $D^{(N_b)}$ with $1\leq N_b\leq N_m$
by minimizing the $E_{N_m}$ at once. 
In other words, we perform variational calculation with 
a set of all variational parameters, $D^{(N_b)}$, 
simultaneously to minimize the $E_{N_m}$.
Its initial state is generated by the refinement process 
in order to save the amount of computation time.
In principle, the FCG yields the closest energy to the exact one within 
a fixed number of basis states, $N_m$, and the refinement process also 
reaches the energy provided by the FCG with a large number of iterations. 
The refinement process needs a far smaller amount of time than that in the FCG, 
and provides us with an approximation good enough to the solution in the FCG.

\begin{wrapfigure}{r}{6.6cm}
  \includegraphics[scale=0.37]{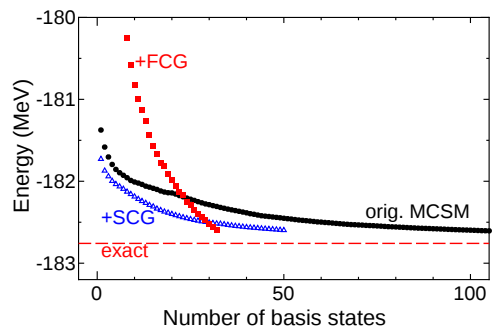}
  \caption{ Convergence patterns of the ground-state 
    energy of $^{72}$Ge in the $f_5pg_9$ shell. The circles, open triangles, 
    and squares denote the results of the original MCSM, SCG, and FCG 
    with $N_m=32$, respectively.
  }
  \label{ge72_dim_cg}
\end{wrapfigure}

Figure  \ref{ge72_dim_cg} shows the convergence 
pattern of the ground-state energy 
of $^{72}$Ge in $f_5pg_9$ shell, which consists of 
the $0f_{7/2}$, $1p_{3/2}$, $1p_{1/2}$, and $0g_{9/2}$ orbits.
Its $M$-scheme dimension is very large and amounts to 140,050,484.
However, it can be handled by the recent shell-model 
diagonalization code \cite{mshell64}.
The energy eigenvalue $ E_{N_b} = \langle \Psi_{N_b} |H|\Psi_{N_b}\rangle $ is plotted 
against the number of basis states, $N_b$.
The SCG method 
attains faster convergence than the original MCSM: 
the SCG gives the same energy as the original MCSM 
in almost half the number of basis dimension.
% provides us with almost the same energy 
% as that provided by the original MCSM
% in roughly half the number of basis states, which means a better convergence. 
The SCG method enables us to compute the energy variance 
in a smaller amount of the computation time than that of the original MCSM, 
since the time to compute the energy variance is proportional to $N_m(N_m+1)/2$.

In the case of the FCG, the number of basis states of the 
variational wave function 
is fixed to $N_m=32$ and we plot energy expectation values in the subspace 
spanned by the $N_b$ basis states. 
In the FCG all basis states, $D^{(1)}, D^{(2)}, ..., $ and $ D^{(32)}$, are optimized 
to minimize $E_{32}$,  
while in the SCG wave function 
$D^{(1)}$ is optimized to minimize  $E_1$. 
Thus, the calculated $E_1$ of the FCG is much higher than the $E_1$ of the SCG and 
the $E_1$ of the original MCSM. 
However, $E_{32}$ of the FCG is lower than that of the SCG and the original MCSM, 
which means the FCG provides us with the best approximation for a fixed 
number of basis states. 
Therefore, we should discuss the energy convergence 
as a function of $N_m$ in the FCG, not as a function of $N_b$, 
unlike the case of the SCG. 

While the FCG yields the best energy with the fixed $N_m$, 
the FCG iteration needs much computational resources 
since we have to replace all basis states at every FCG iteration.
Which level of the optimization is most efficient in terms 
of the computation time 
depends on the property of the objective wave function. 
For applications, we adopt the original MCSM method 
and the SCG method in Sect. \ref{sect3_tabe}
and the refinement process in Sect. \ref{sect4_ytsunoda}.

\subsection{Energy-variance extrapolation }
\label{sec:eve}

Even if the variational procedure works excellently, 
a small gap between the exact energy eigenvalue 
and the energy expectation value 
of the approximated wave function remains. 
The gap can be removed by extrapolation procedures, 
which have been studied 
intensively \cite{horoi_conv, extrap_2ndorder, papenbrock_factorize, shen_zhao}. 
Among them, the energy-variance extrapolation method 
is a general framework for the supplementation of the variational 
calculation and is rather independent of the form of the approximated wave function.
This method
has been known in condensed matter physics \cite{imada_pirg} 
and was firstly introduced in shell-model calculations 
with conventional particle-hole truncation 
by T. Mizusaki and M. Imada \cite{extrap_2ndorder}. 
In the present work, we apply this extrapolation method
to estimate the exact eigenvalue precisely
by utilizing a sequence of the 
linear combinations of the projected Slater determinants, $|\Psi_{N_b}\rangle$ with 
$1\leq N_b \leq N_m$, which have been obtained in the variational procedure
discussed in Sect.\ref{sec:var}.
The major obstacle of the energy-variance extrapolation
is the necessity of the computation of 
the energy variance. Its efficient computation
is described in Sect.\ref{sec:computational_aspects}.

The energy-variance extrapolation method
is based on the fact that the energy variance of the 
exact eigenstate is zero. 
The energy variance of the approximated wave function 
is not exactly zero, but rather small and approaches 
zero as the approximation is improved.
In the framework of the energy-variance extrapolation, 
we draw the energy $E_{N_b} = \langle \Psi_{N_b} | H |\Psi_{N_b} \rangle $ 
against the energy variance 
$\langle \Delta H^2 \rangle_{N_b} = \langle \Psi_{N_b} | H^2 |\Psi_{N_b} \rangle - E_{N_b}^2 $, 
the plot of which is called ``energy-variance plot''.
The variance usually approaches zero as $N_b$ increases,  
as the point in the energy-variance plot approaches the $y$-intercept. 
Following the idea of Ref.\citen{extrap_2ndorder}, 
These points are usually fitted by a first- or second-order polynomial such as 
\begin{equation}
  E = c_0 + c_1 \langle \Delta H^2 \rangle  + c_2 (\langle \Delta H^2 \rangle )^2, 
\end{equation}
where these coefficients $c_0$, $c_1$, and $c_2$ are determined by least square fit. 
By extrapolating the fitted curve into the $y$-intercept we obtain the 
extrapolated energy, namely, $c_0$.

In the framework of the present study, 
the variational procedure discussed in Sect. \ref{sec:var} 
provides us with a sequence of approximated 
wave functions, which can be utilized in the energy-variance extrapolation method. 
The SCG procedure provides us with a successive
sequence of the wave functions, $|\Psi_{N_b}\rangle$ 
with $1\leq N_b \leq N_m$, 
where $N_m$ is the maximum of $N_b$.
For each $N_b$, we evaluate the energy $E_{N_b}$ and 
energy variance $\langle \Delta H^2 \rangle_{N_b} $. 
Here, we demonstrate how the extrapolation method works 
with $^{56}$Ni in the $pf$ shell as an example.
The effective interaction is the FPD6 interaction \cite{fpd6}, 
which was adopted also in Ref. \citen{mcsm_extrap}. 

\begin{figure}[htbp]
  \begin{center}
  \includegraphics[scale=0.4]{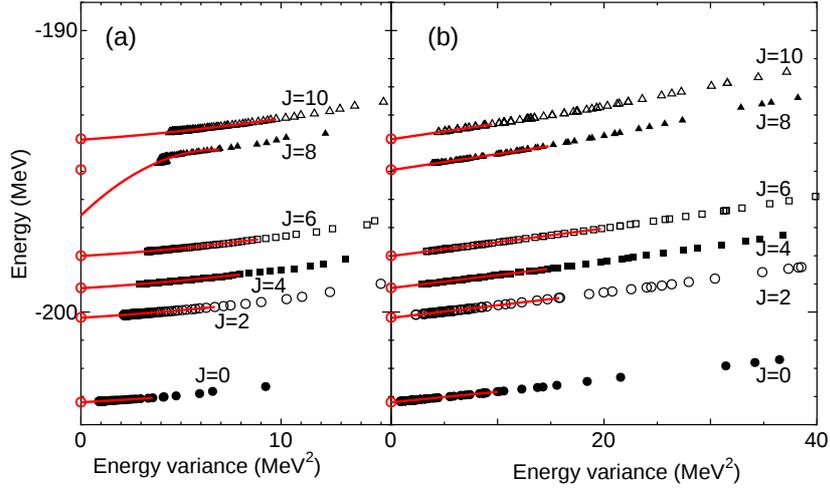}
  \end{center}
  \caption{ Energy-variance plot for $^{56}$Ni 
    (a) without reordering, (b) with reordering. 
    The open circles on the $y$-axis denote the exact shell-model energies. 
    See text for details. }
  \label{ni56_yrast}
\end{figure}

Figure \ref{ni56_yrast}(a) shows the energy-variance plot 
of the SCG method for yrast states of $^{56}$Ni. 
In this calculation, we take the $K=I$ state only in the angular-momentum 
projector, namely $f_{n,K}^{(N_b)}=0$ if $K\neq I$ in Eq.(\ref{eq:wf}), 
for simplicity. 
As $N_b$ increases the energy-variance point moves smoothly and 
approaches the $y$-axis or variance zero, except for the $8^+$ state. 
The fitted curves for these points are shown as red solid lines, 
and they also show smooth behavior. 
The extrapolated energies, or $y$-intercepts of the fitted curves, 
agree quite well with the exact ones, which are shown as open circles 
on the $y$-axis.
Especially concerning the ground-state energy, 
the minimum variance of the approximated wave function 
is $\langle \Delta H^2 \rangle_{N_b=100} = 0.89$MeV$^2$,  
which is smaller and gives a better approximation than the 
result of the original MCSM, 
$\langle \Delta H^2 \rangle_{N_b=150} = 1.05$MeV$^2$ \cite{mcsm_extrap}, 
mainly thanks to the introduction of the CG method.

On the other hand, the plot of the $8^+$ state shows the anomalous behavior, 
in which the energy decreases when increasing the number of basis states 
but the variance does not decrease 
at $\langle \Delta H^2 \rangle \simeq 4$MeV$^2$.
As a result, the simple extrapolation method apparently fails.
A straightforward solution to this failure is to increase $N_b$ 
until the extrapolation method works, and it is shown in Ref. \citen{okinawa_shimizu}. 
However, in other cases, it might be difficult 
due to the increase of computation time. 
We discuss another remedy for such cases in the following section.

\subsection{Reordering technique to improve the energy-variance
extrapolation}
\label{sec:eve_ni56}

% order is determined by variational proces
% in priciple it can change
% choose best order

The anomalous behavior can be removed by 
the reordering of the basis states \cite{shimizu_reordering}. 
In this section, we demonstrate that the reordering technique
makes the energy-variance extrapolation stable 
and avoids the difficulty of an anomalous kink 
such as the $8^+_1$ state of $^{56}$Ni discussed in the previous
subsection.

A sequence of the approximated wave functions, 
($|\Psi_1\rangle$, $|\Psi_2\rangle$, ..., $|\Psi_{N_b}\rangle$, ..., $|\Psi_{N_m}\rangle$)
is specified by a set of basis states and its order
($|\phi_1\rangle$, $|\phi_2\rangle$, ..., $|\phi_{N_b}\rangle$, ..., $|\phi_{N_m}\rangle$).
In the SCG method, the order of the basis states is determined by 
the variational procedure. 
However, we can shuffle the order of basis states and 
make another sequence of the approximated wave functions without 
additional computational effort. 
If we assume that 
the extrapolated value is independent of the order of the basis states, 
there exists the best order which makes the extrapolation procedure stable. 
In the reordering method, the order is determined 
so that the gradient of the curve in the energy-variance plot 
is as small as possible. 
The practical algorithm to determine the 
order is discussed in Ref. \citen{shimizu_reordering}.

\begin{wrapfigure}{r}{6.6cm}
  \includegraphics[scale=0.35]{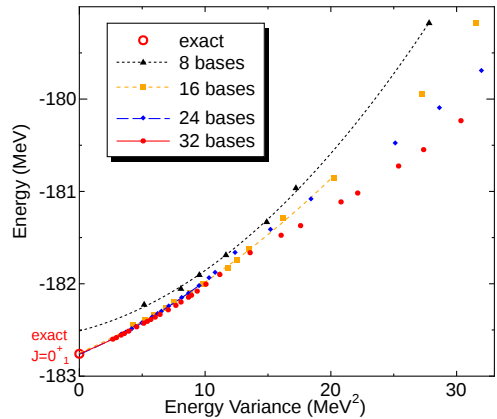}
  \caption{Energy vs. energy variance plot 
    of the ground state energy  
    of $^{72}$Ge in the $f_5pg_9$ shell. 
    The wave functions are provided by 
    the FCG with $N_m=8$ (triangles), 16 (squares),
    24 (diamonds), and 32 (circles). 
}
  \label{ge72_mcg}
\end{wrapfigure}

Figure \ref{ni56_yrast}(b) shows the extrapolation plot 
with the reordering technique 
using the same set of the basis states as in Fig. \ref{ni56_yrast}(a).
In this case, the gradient of the fitted curve is so stable
that the points in the energy-variance plot are fitted by a first-order polynomial 
and the region to be used for the fit is rather broad.
In Fig. \ref{ni56_yrast}(b), 
the anomalous behavior of the variance plot of the $8^+$ state 
disappears and the extrapolated value agrees with the exact energy quite well.
$^{56}$Ni is also known to have a shape coexistence feature \cite{ni_coex_mcsm}, 
which is plausible to be the origin of the anomalous behavior of the $8^+$ state 
in Fig. \ref{ni56_yrast}(a) 
like the case of $^{72}$Ge discussed in Ref.\citen{shimizu_reordering}.

We demonstrate another example of the energy-variance extrapolation method combined 
with the reordering technique 
in Fig. \ref{ge72_mcg} using the FCG wave functions of 
$^{72}$Ge with the JUN45 effective interaction \cite{jun45}.
In this figure, the energy expectation value of the FCG
wave function is plotted as a function of the corresponding energy variance
with the basis states being 
$N_m=8, 16, 24,$ and $32$, and their second-order fitted curves.
The order of the basis states in each sequence is determined 
by the reordering technique\cite{shimizu_reordering}.
Note that there is no specific order in the FCG procedure 
because we treat all basis states on an equal footing.
The extrapolated energy apparently converges except for $N_m=8$. 
While we show the second-order fit in the figure,
even the extrapolated value of the first-order fit 
agrees well with that of the second-order fit in the case of $N_m=32$.

\subsection{Computational aspects}
\label{sec:computational_aspects}

The large-scale shell-model calculation is one of the challenging issues 
for nuclear physics. It is essential to develop 
a program which runs efficiently on recent supercomputers. 
Concerning a calculation using a single CPU, 
we proposed a sophisticated way of efficient computation 
of the matrix element of non-orthogonal Slater determinants 
in Ref.\citen{mcsm_tuning}. 
Moreover, since the main trend of recent supercomputers favors massively parallel 
computers, the parallel efficiency is worth discussing for future studies.

In the case of the SCG method, we need to compute 
the Hamiltonian matrix elements 
and the gradient vector concerning $D^{(n)}$ of 
two angular-momentum-, parity-projected Slater determinants.
The three-dimensional integral 
of the Euler angles in Eq.(\ref{eq:proj})
is performed by discretizing each range of the integral 
into mesh points using the Gauss-Chebyshev quadrature for 
the $z$-axis rotations and the Gauss-Legendre quadrature for the $y$-axis rotation, 
which is shown in Eq.(\ref{eq:proj_sum}). 
The numbers of the mesh points are taken as 26 for the $z$-axis rotation 
and 16 for the $y$-axis rotation for example. 
The parity projection is equivalent to two mesh points.
The product of these mesh points, which are denoted 
by $\lambda$ in Eq.(\ref{eq:proj_sum}) give rise to 
$N_{\rm mesh} = 26^2\times 16\times 2= 21632$ components 
in evaluating a matrix element 
in terms of the projected Slater determinants. 
Since these components can be computed independently, 
the program was written for massive parallel computation.
When we apply the matrix-product technique discussed 
in Ref.\citen{mcsm_tuning} 
with the bunch size $N_{\rm bunch}$ being e.g. 30,  
we still have the $N_b N_{\rm mesh}/N_{\rm bunch} \simeq 721N_b$ elements 
to be computed in parallel. 

\begin{figure}[htbp]
  \includegraphics[scale=0.4]{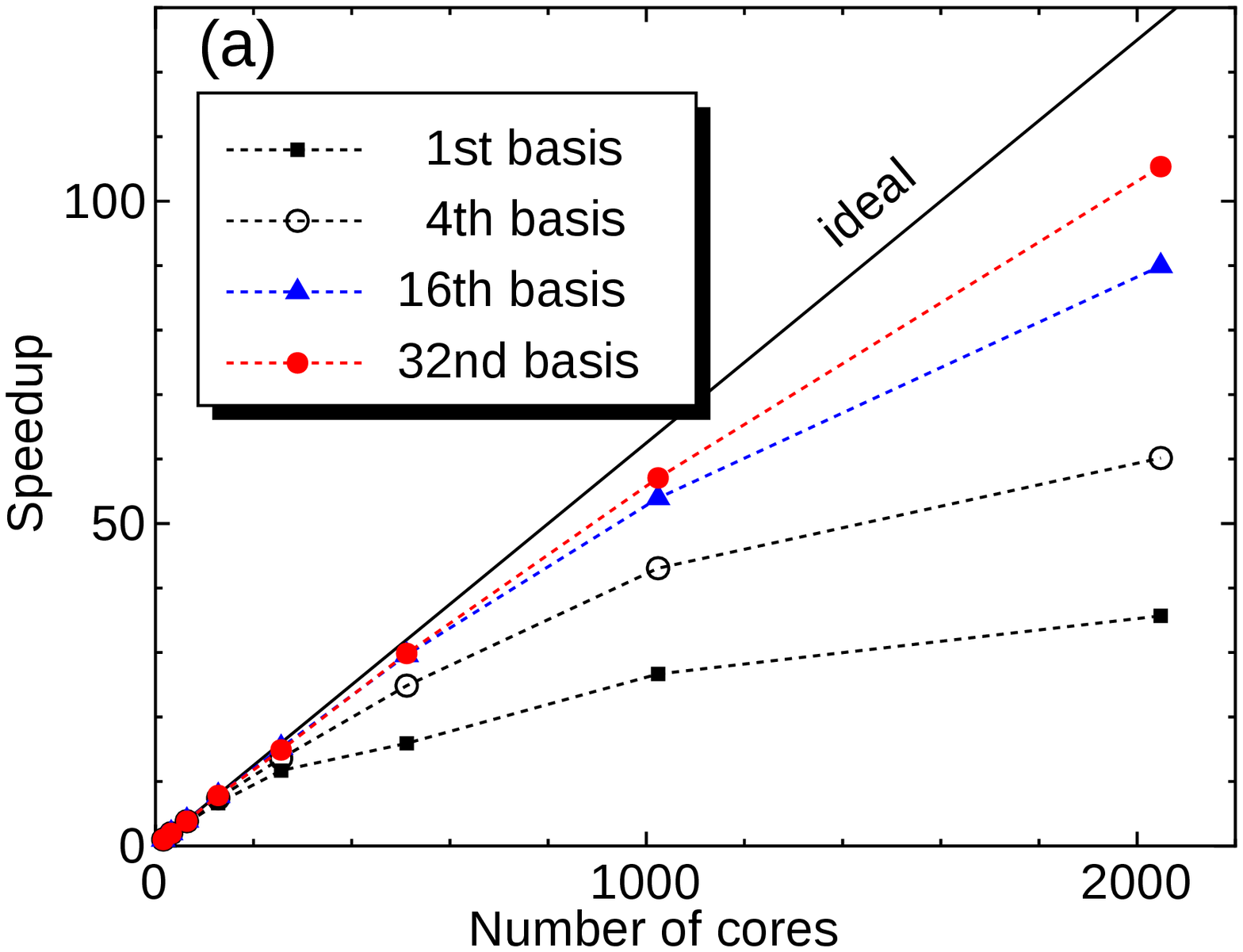}
  \includegraphics[scale=0.4]{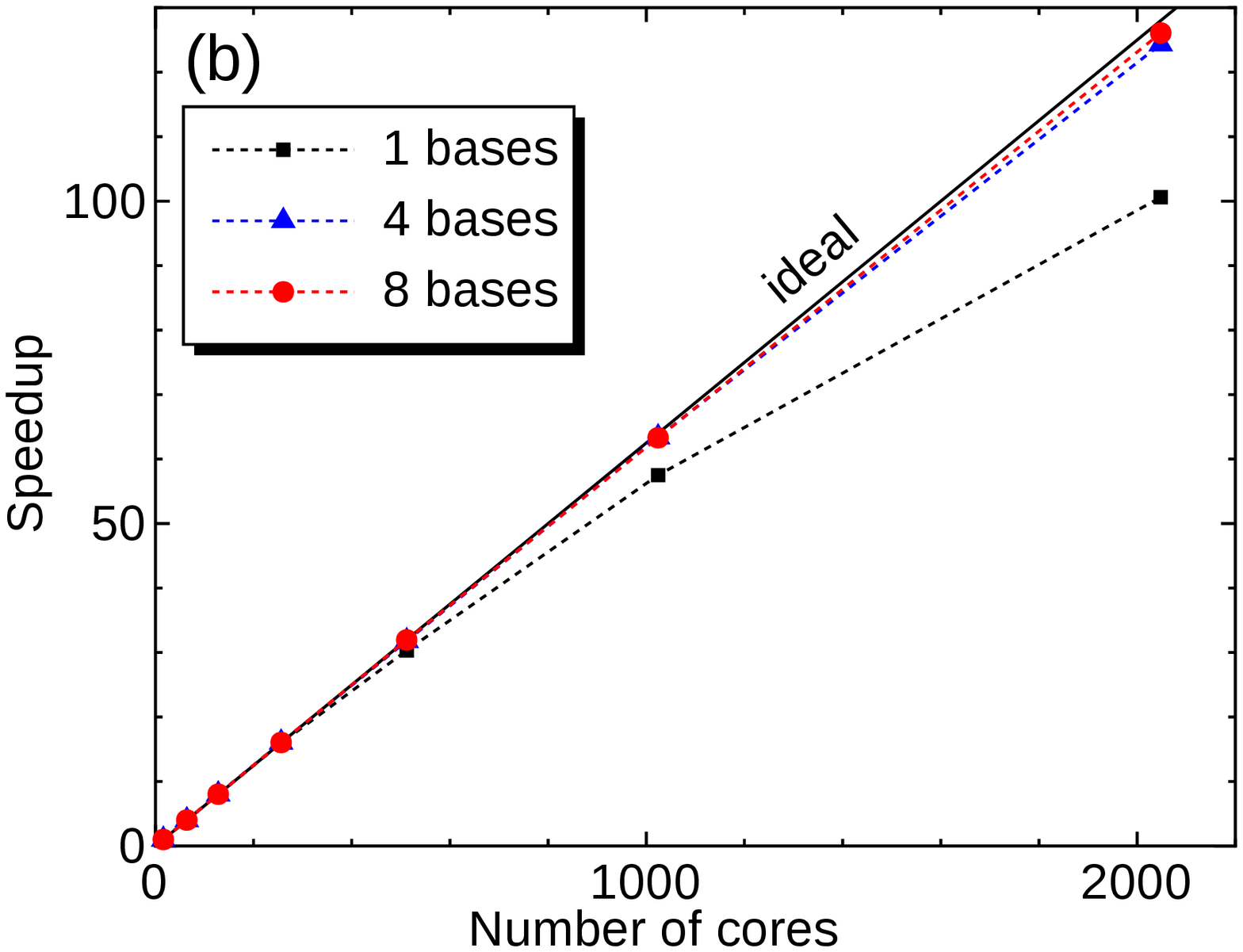}
  \caption{(a) Speedup of the parallel computation 
    of the SCG process in unit of 
    the computation time using 16 CPU cores.
    The squares, open circles, triangles and filled circles 
    represent 
    the inverses of computation times of the variational process 
    to obtain 1st, 4th, 16th, and 32nd 
    basis states respectively.
    The solid line shows ideal scaling to guide the eyes. 
    (b) Speedup of the parallel computation of the energy variance. 
    The squares, triangles, and circles represent 
    the inverses of computation times of 1, 4, and 
    8 basis states, respectively.
  }
  \label{ge64_parallel}
\end{figure}

Figure \ref{ge64_parallel} shows
the parallel efficiency of the benchmark calculation 
of the ground state of $^{64}$Ge  as an example. 
The model space consists of the $pf$ shell and $g_{9/2}$ orbit
and the PFG9B3 effective interaction is used \cite{pfg9b3}. 
Its  $M$-scheme dimension reaches $1.7\times 10^{14}$, which is far 
beyond the current limitation of the Lanczos method.
The MCSM result of this system was already reported 
in Refs. \citen{mcsm_extrap, shimizu_reordering}.
This benchmark was performed using the Intel Fortran compiler ver.11.0 
\cite{intel_mkl}
on the T2K open Supercomputer at the University of Tokyo \cite{t2k-todai}.

Figure \ref{ge64_parallel} (a) shows 
the performance gain of the parallel computation 
of the SCG process to determine the  1st, 4th, 8th, 16th, and 32nd 
basis states respectively. 
Although the parallel efficiency for calculating the first basis state is 
not good because of the small amount of computation, 
the efficiency  for the 32nd basis state 
with 2048 CPU cores reaches 82\% of that with 16 cores.

We calculate the energy variance using the formula shown in 
Appx.\ref{sec:eqs}\cite{mcsm_extrap}. 
Because the two-body matrix elements in the $M$-scheme, $v_{ijkl}$, 
are sparse due to the symmetry which the Hamiltonian has, 
we store in memory only non-zero matrix elements in block-diagonal form 
by treating $v_{(ij), (kl)}$ as a rank-2 matrix  with indices 
$(ij)$ and $(kl)$. 
Thus, we can compute the energy variance efficiently, 
and the detail of the practical computation is written in Appx.\ref{sec:eqs}.
In a similar manner to the case of the variational process, 
we compute the energy variance by dividing the whole computation 
into matrix elements 
which are moreover divided into each mesh point of Eq.(\ref{eq:proj}), 
resulting in $N_b(N_b+1) \times N_{\rm mesh}$ independent components 
to be computed in parallel.
In addition, we do not need an iterative process like the CG method,  
and therefore 
a small amount of network communication appears only at the beginning 
and at the end of the computation.
Thus the performance scaling of the parallel computation seems perfect 
at $N_b \geq 4$, 
which is shown in Fig. \ref{ge64_parallel}(b).

In practice, it took totally 807 seconds to obtain 
an SCG wave function of $^{64}$Ge $0^+_1$ state with 32 basis states, 
and it took 588 seconds to compute the energy variance of this SCG 
wave function using 2048 CPU cores.

%%%% Sect.3 no-core calc. T. Abe
%\input{sect3_tabe.tex}
% \newpage

%------------------------------------------------------------------------------
\section{Application of the MCSM to the {\it ab initio} shell model}
\label{sect3_tabe}
%------------------------------------------------------------------------------

In this section we focus on the latest application of the MCSM to the {\it ab initio} shell model calculations, which has become feasible recently with the aid of major development of the MCSM algorithm discussed in Sect.~\ref{sect2_shimizu} and also a remarkable growth in the computational power of state-of-the-art supercomputers. 
First, the no-core shell model (NCSM) and its variants are briefly reviewed. 
The limitation of the NCSM and the motivation for the application of the MCSM to the {\it ab initio} no-core Full Configuration Interaction (FCI) approach are further discussed here. 
Then, the current status of the benchmarks in the no-core MCSM is referred to based on the results mostly from Ref.~\cite{Abe:2012wp}. 
Finally our challenge to visualize the intrinsic states constructed by superpositioned non-orthogonal Slater determinants is also demonstrated. 

%------------------------------------------------------------------------------
\subsection{{\it Ab initio} shell models}

One of the major challenges in nuclear theory is to understand nuclear structure and reactions from {\it ab initio} methods. 
{\it Ab initio} calculations for nuclear many-body systems beyond $A = 4$ have 
recently become feasible due to the rapid evolution of 
computational technologies these days. 
In {\it ab initio} approaches for the nuclear structure calculations, all the nucleons constituting the nucleus are considered as the fundamental degrees of freedom and the bare/effective interactions based on realistic nuclear forces are adopted. 
As for bare two- and three-nucleon interactions, 
the phase-shift equivalent family of two-nucleon 
interactions, derived from the meson-exchange theory and 
chiral Effective Field Theory, 
in addition to three-nucleon interactions 
\cite{Epelbaum,N3LO,Wiringa:1994wb,Pieper_3NF,Illinois} is usually used. 

{\it Ab initio} NCSM has been emerging for about a decade and is now available 
for the study of nuclear structure and reactions in the $p$-shell nuclei \cite{NCSM}. 
Unlike the conventional shell model with a core, 
the NCSM does not assume an inert core just like the name itself implies and treats all the nucleons composing the nucleus on an equal footing. The NCSM is thus said to be one of the {\it ab initio} approaches 
along with the Green's Function Monte Carlo % (GFMC) 
\cite{GFMC} and Coupled Cluster % (CC) 
theory \cite{CC}. 
In the NCSM (in a narrow sense)\cite{NCSM} the model (or basis) space is usually truncated 
by the so-called $N_{max}$, which is the sum of the excitation quanta
 above the reference state. 
The effective interactions renormalized to that model space are used so as to obtain the faster convergences of the energy with respect to $N_{max}$. 
Generally, the effective interactions are derived 
by the so-called Lee-Suzuki-Okamoto method \cite{Lee_Suzuki_Okamoto}. 
The NCSM result approaches the exact solution either by taking the larger model space with the level of the cluster approximation fixed or by improving the order of the cluster expansion with the model space fixed. 

A similar but distinct approach to the NCSM is the No-Core Full Configuration (NCFC) approach \cite{NCFC}. 
The NCSM result by using the effective interactions derived 
by the Lee-Suzuki-Okamoto procedure approaches the exact solution either 
from below or above due to the violation of the strict variational upper bound 
of the exact solution. 
Therefore the extrapolation of the NCSM result into the infinite model space is obscure. 
The NCFC method employs the bare or effective (softened) low-momentum interactions evolved from bare nuclear forces by the renormalization group transformations \cite{RG}, which validates the variational upper bound of the calculated energy. The NCFC enables access to full {\it ab initio} solutions by a simple extrapolation into the infinite model space in the two-dimensional parameter space ($\hbar \omega$, $N_{max}$). 
One of the advantages both in the NCSM and NCFC methods is the perfect factorization of the intrinsic and the center-of-mass wave functions, so that the intrinsic state does not suffer from the spurious center-of-mass motion. 

\begin{wrapfigure}{r}{6.6cm}
% \begin{figure}
\centerline{
\includegraphics[scale=1.2]{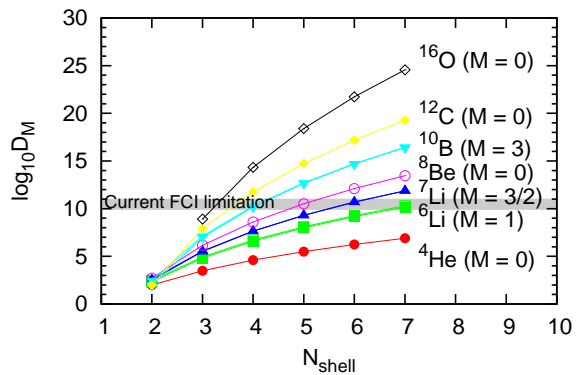}}
\caption{$M$-scheme dimensions as functions of basis-space size, $N_{shell}$.}
\label{Fig.Mdim_Nshell}
% \end{figure}
\end{wrapfigure}

As {\it ab initio} approaches treat all of the nucleons democratically, computational demands for the calculations explode exponentially as the number of nucleons and/or the model spaces increase. 
Current limitation of the direct diagonalization of the Hamiltonian matrix by the Lanczos iteration is around the order of $10^{10}$ shown in Fig.~\ref{Fig.Mdim_Nshell}. 
So far the largest calculations have been done in the $^{14}$N with $N_{max} = 8$ which results in the $M$-scheme many-body matrix dimensions being $\sim 10^{9}$ and associated non-vanishing 
three-nucleon force matrix elements being $\sim 4 \times 10^{13}$ \cite{Maris:2011as}. 
In order to access heavier nuclei beyond the $p$-shell region with larger model spaces by {\it ab initio} shell-model methods, many efforts have been devoted for several years. 
One of these approaches in the $N_{max}$ truncation is the Importance-Truncated NCSM (IT-NCSM) \cite{Roth}. 
In the IT-NCSM, the model spaces are extended by using the importance measure 
evaluated by the perturbation theory. 
Another approach is the Symmetry-Adapted NCSM (SA-NCSM) \cite{Dytrych}, 
where the model spaces are truncated by the selected symmetry groups. 

%------------------------------------------------------------------------------
%------------------------------------------------------------------------------

Besides the $N_{max}$ truncation of the model space in the {\it ab initio} shell models, there is the FCI method to give the exact solutions in the fixed model space. 
Different from the $N_{max}$ truncation in the NCSM and NCFC methods, the FCI truncates the model space by the single-particle states, so-called $N_{shell}$ or $e_{max} (\equiv N_{shell} - 1)$. 
% In principle the FCI in $N_{shell} \rightarrow \infty$ and the NCSM in $N_{max} \rightarrow \infty$ with the fixed cluster approximation, $a$, or in the fixed $N_{max}$ with $a \rightarrow A$ should agree with the NCFC, which gives the extrapolated solutions into $N_{max} \rightarrow \infty$. 
% The full solutions among them, however, have not been examined yet. 
As shown in Fig.~\ref{Fig.Mdim_Nshell}, the explosion of the dimensionality prohibits the full {\it ab initio} solutions of the FCI (and also the NCSM) beyond the lower p-shell region. 
Similar to the attempts of the IT-NCSM and SA-NCSM, the MCSM is one of the promising candidates to go beyond the FCI method \cite{okinawa_tabe, Abe:2012wp}. 
Note that there is a similar approach to the no-core MCSM 
referred to as the Hybrid Multi-Determinant method \cite{Puddu}. 
In the following subsection we will show some recent investigations by the {\it ab initio} no-core MCSM.

%------------------------------------------------------------------------------
\subsection{Benchmarks of the MCSM to the {\it ab initio} no-core FCI}
%------------------------------------------------------------------------------

As an exploratory work of the original % existing (original) 
MCSM has been applied to the no-core calculations for the structure and spectroscopy of the beryllium isotopes \cite{Liu2011}. 
In Ref.~\cite{Liu2011} the low-lying excited states of $^{10}$Be and $^{12}$Be are investigated. The excitation energies of the first and second $2^+$ states and the B(E2; 2$^+_{1}\rightarrow$ 0$^+_{g.s.}$) for $^{10}$Be with a treatment of spurious center-of-mass motion show good agreement with experimental data. The deformation properties of the $2^+_1$ and $2^+_2$ states for $^{10}$Be and of the $2^+_1$ state for $^{12}$Be are studied in terms of electric quadrupole moments, E2 transitions and the single-particle occupations. 
The triaxial deformation of $^{10}$Be is also discussed in terms of the B(E2; 2$^+_{2}\rightarrow$ 2$^+_{1}$) value. 
This work motivates a further extension of the MCSM application to the {\it ab initio} FCI calculations \cite{okinawa_tabe}. 
Currently, the availability of the MCSM for the no-core calculations has been tested extensively in light nuclei \cite{Abe:2012wp}. 

%------------------------------------------------------------------------------
\begin{figure}[htbp]
%\begin{center}
\begin{tabular}{cc}
\resizebox{67.5mm}{!}{\includegraphics{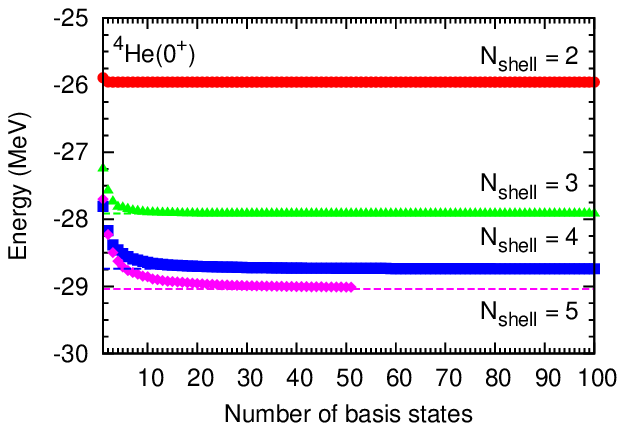}} & 
\resizebox{67.5mm}{!}{\includegraphics{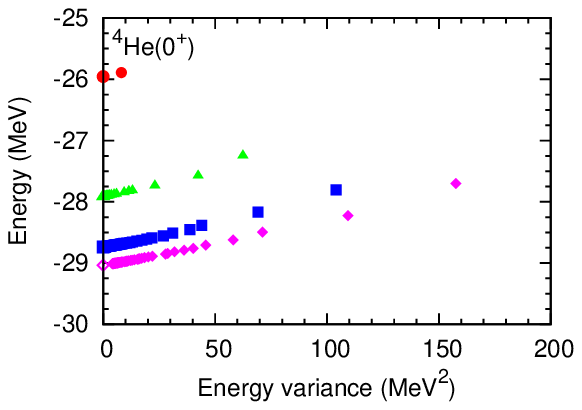}} \\
\end{tabular}
\caption{
  $^4$He ground-state energies as functions of number of basis states (left) 
  and energy variance (right). 
The red, green, blue and purple solid symbols (horizontal dashed lines in the left figure and open symbols at the zero energy variance in the right figure) are the MCSM (FCI) results in $N_{shell} = 2$, $3$, $4$ and $5$, respectively. The harmonic oscillator energies are taken at optimal values for each state and model space. The Coulomb interaction and the spurious center-of-mass motion effect are not considered. Isospin symmetry is assumed.
}
\label{Fig.E_4He_NCMCSM}
%\end{center}
\end{figure}
%------------------------------------------------------------------------------

%------------------------------------------------------------------------------
\begin{figure}
\centerline{
\includegraphics[width=\columnwidth]{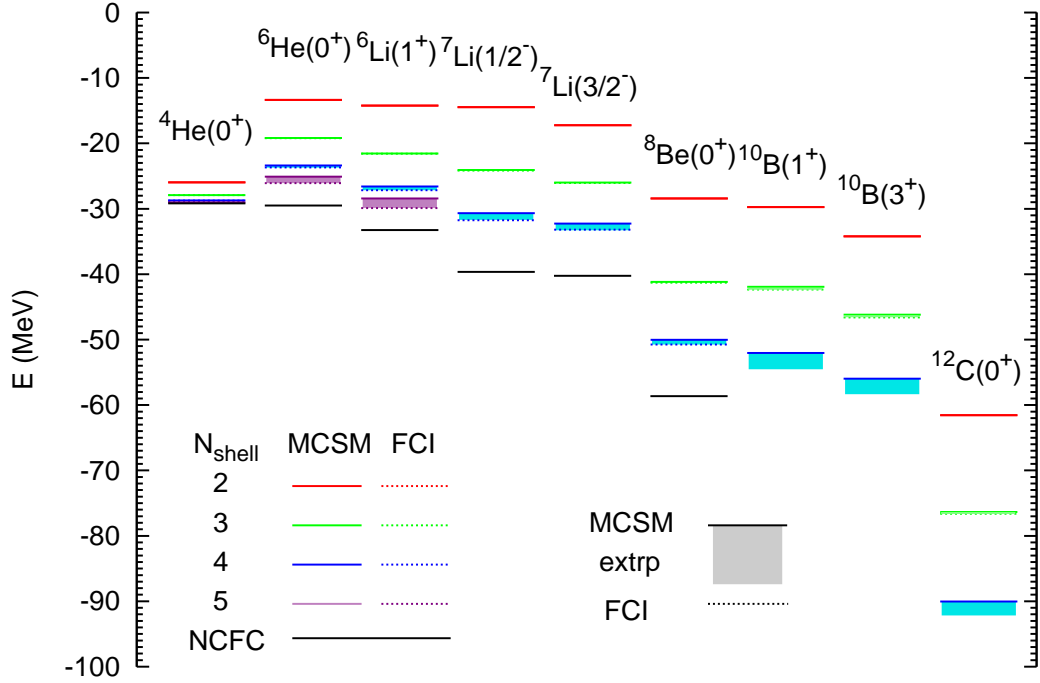}}
\caption{
  Comparisons of the energies between the MCSM and FCI along
  with the fully converged NCFC results where available \cite{Abe:2012wp}. The MCSM
  (FCI) results are shown as the solid (dashed) lines that nearly
  coincide where both are available.  The extrapolated MCSM results
  are illustrated by bands.  From top to bottom, the truncation of the
  model space is $N_{shell} = 2$ (red), $3$ (green), $4$ (blue) and
  $5$ (purple).  Note that the MCSM results are extrapolated by the
  energy variance with the second-order polynomials \cite{mcsm_extrap}.  
  Also note that all of the results of $^{10}$B and $^{12}$C at 
  $N_{shell} = 4$ were obtained only with MCSM.}
  \label{Fig:E_all_NCMCSM}
\end{figure}
%------------------------------------------------------------------------------

As a typical example, the behavior of the ground-state energies of $^4$He ($0^+$) 
with respect to the number of basis states 
and to the energy variance in $N_{shell} = 2 - 5$ are shown in Fig. \ref{Fig.E_4He_NCMCSM}. 
Figure \ref{Fig:E_all_NCMCSM} illustrates the comparisons of the energies for each state and model space between the MCSM and FCI methods. 
The FCI gives the exact energies in the fixed size of the mode space, 
while the MCSM gives approximated ones. Thus the comparisons between them show how well the MCSM works in no-core calculations. 
For this benchmark comparison, the JISP16 two-nucleon interaction 
is adopted and the Coulomb force is turned off. Isospin symmetry is assumed. 
The energies are evaluated for the optimal harmonic oscillator frequencies where the calculated energies are minimized for each state and model space. 
Here the contributions from the spurious center-of-mass motion are ignored for simplicity. 
In Fig.~\ref{Fig:E_all_NCMCSM}, the comparisons are made for the states; $^4$He($0^+$), $^6$He($0^+$), $^6$Li($1^+$), $^7$Li($1/2^-$, $3/2^-$), $^8$Be($0^+$), $^{10}$B($1^+$, $3^+$) and $^{12}$C($0^+$). 
The model space ranges from $N_{shell} = 2$ through $5$ for $A \le 6$ ($4$ for $A \ge 7$). 
Note that the energies of $^{10}$B($1^+$, $3^+$) and $^{12}$C($0^+$) in $N_{shell} = 4$ 
are available only from the MCSM results. The $M$-scheme dimensions for these states ($1.82 \times 10^{10}$ for $M = 1$ and $1.52 \times 10^{10}$ for $M = 3$ in $^{10}$B and $5.87 \times 10^{11}$ for $M = 0$ in $^{12}$C) are already marginal or exceed the current limitation in the FCI approach. 
The number of basis states  are taken up to $100$ in 
$N_{shell} = 2  - 4$ and $50$ in $N_{shell} = 5$. 
In Fig.~\ref{Fig:E_all_NCMCSM}, the solid (dashed) lines indicate the MCSM (FCI) results.
The shaded regions express the extrapolations in the MCSM, and 
the lower bound of the shaded region corresponds to the extrapolated energy.
Furthermore, we also plot the NCFC results for the states of $4 \le A \le 8$ 
as the fully converged energies in the infinite model space. 
As seen in Fig.~\ref{Fig:E_all_NCMCSM}, the energies are consistent with
each other where the FCI results are available to 
within $\sim 100$ keV ($\sim 500$ keV) at most of the MCSM results with(out) the 
energy-variance extrapolation in the MCSM. 
The other observables besides the energies also give reasonable agreements between the MCSM and FCI results. 
The detailed comparisons among the MCSM, FCI and NCFC methods can be found in Ref.~\citen{Abe:2012wp}. 

By exploiting the recent development in the computation of 
the Hamiltonian matrix elements 
between non-orthogonal Slater determinants \cite{Utsuno:2012vm} 
and the technique of energy-variance extrapolation 
\cite{mcsm_extrap}, 
the observables give good agreement between the MCSM and FCI results in the $p$-shell nuclei. 
From the benchmark comparison, the no-core MCSM is now verified in the application 
to the {\it ab initio} no-core calculations for light nuclei. 
Moreover the application of the no-core MCSM to heavier nuclei is expected in the near future.

% moved the outlook

%Following the ealier attempt \cite{Liu2011} and the benchmark calculations of the no-core MCSM \cite{okinawa_tabe, Abe:2012wp}, further confirmations for the treatment of the spurious center-of-motion effects and the inclusion of the Coulomb force (and higher electromagnetic effects) are needed for providing the actual comparison with the experimental data. 

%After these confirmations, the no-core MCSM is expected to extend the calculations in larger model spaces for the $sd$-shell nuclei, heavier nuclei beyond $p$-shell nuclei. 
%Concerning to the further extensions in the MCSM algorithm, 
%the inclusion of the 3-body interaction is inevitable to understand the nature of 3N forces in the nuclear structure. 
%Moreover the coupling to the continuum state is absolutely essential to apply the no-core MCSM to the neutron-rich neuclei far beyond the stablity line. 

%The reproduction and visualization of the cluster-like state appeared near the threshhold in the light nuclei is also a fascinating topic in the framework of the nuclear shell model. Further investigations to challenge the cluster states by the no-core MCSM are necessary in near future. 

%%%% Sect.3.2 no-core calc. intrinsic state T. Yoshida
%\input{sect32_yoshida.tex}
\subsection{Analysis of intrinsic state}
While {\it ab initio} approaches have been studied intensively in light nuclei, 
it is relatively difficult to study the cluster structure in an {\it ab initio} way.
Among these approaches, 
the Green's Function Monte Carlo first provided the two-$\alpha$ structure 
of the $^8$Be ground state illustratively\cite{yoshida:2000wp}.
This study has shown the possibility
 that the cluster structure can appear in $^8$Be,
without assuming any cluster structure in advance.
Generalizing this result, 
it may be possible to treat cluster structure 
 from a pure single particle picture. % when the model space is large enough.
In this subsection,  we show how to visualize the cluster state 
in the no-core MCSM calculation and by analyzing the calculations 
we discuss the appearance of $\alpha$ cluster structure.
% The calculation contributes to the
%  improvement of the description around the cluster threshold region by the effect of cluster configuration.
%
It is also suited to clarify the relation between the shell-model 
and cluster pictures \cite{yoshida:2005im}
from the shell-model point of view.
This view point has not been investigated very well yet.
%
%%%%%%%%%%%%%%%%%%%%%%%%%%%%%%%%%%%%%%%%
% ADDED 20120427
%%%%%%%%%%%%%%%%%%%%%%%%%%%%%%%%%%%%%%%%
%<<<<<<<<<<<<<
Recently, the density profile in the lithium isotopes has been investigated 
by the NCFC \cite{yoshida:2012NCFC}.
The method has shown how to calculate the translationally-invariant density. 
In Li isotopes, the shape distortion and cluster-like structure has been found.
Thus, the study of cluster structure has become
a realistic subject by using the shell-model calculation.

To extract the cluster structure from the no-core MCSM, we 
define the intrinsic state to 
visualize the cluster shape in the intrinsic framework
which is extracted from the angular-momentum-projected wave function. 
The wave function of the no-core MCSM, which is defined in Eq.(\ref{eq:wf}), 
is represented as an angular-momentum projection of a linear combination of basis states 
such as
\begin{equation}
%   |\Psi \rangle = P^I | \Phi^{\rm (BR)}\rangle  
%   \ \ \ \ \   | \Phi^{\rm (BR)}\rangle = \sum_n  f_n |\phi_n\rangle,
  |\Psi \rangle = P^I | \Phi\rangle  
  \ \ \ \ \   | \Phi\rangle = \sum_n  f_n |\phi_n\rangle,
\end{equation}
where the total $I$ is assumed to be zero and 
$K$-quantum number and parity projections
are omitted for simplicity.
This linear combination of the unprojected basis states, $|\Phi\rangle$, 
% ,  which is referred to as ``before rotation'' (BR), 
cannot be considered  as an intrinsic state
because the principal axis of a basis state, 
$|\phi_i\rangle$, is not in the same direction 
as that of another basis state. 
Therefore we rotate each basis state so that 
it has a diagonalized quadrupole-moment; $Q_{zz}>Q_{yy}>Q_{xx}$ 
and $Q_{ij}=0, (i\ne j)$, respectively, 
following the concept of Ref. \citen{yoshida:2000wp}.
As a result, these rotated basis states have a large overlap with each other 
and make a distinct principal axis toward the $z$-axis. 
The intrinsic wave function $|\Phi^{{\rm intr}} \rangle$ is defined as 
\begin{eqnarray}
  |\Phi^{{\rm intr}} \rangle &\equiv& \sum_{n} f_{n} R(\Omega_n)|\phi_n \rangle = \sum_{n} f_{n}|\phi^R_n\rangle, 
  \label{DENS}
\end{eqnarray}
where the $R(\Omega_n)$ is a rotation operator with Euler's angle $\Omega_n$.
The $\Omega_n$ is determined so that 
the transformed basis state $|\phi_n^R\rangle = R(\Omega_n) | \phi_n\rangle$
has the diagonalized quadrupole-moment.
The transformed coefficient $D^R_n$ (by $R(\Omega_n)$) is derived by the relation 
in Ref. \citen{ppnp_mcsm}.
This state exactly has the same energy after the angular momentum projection. 
We calculate the one-body density of the intrinsic state such as 
\begin{eqnarray}
  \rho^{{\rm intr}} (r)
  &=& \langle \Phi^{{\rm intr}} |\sum_i \delta(r-r_i) |   \Phi^{{\rm intr}} \rangle,
\label{DENS1}
\end{eqnarray}
where $r_i$ denotes the position of the $i$-th nucleon.

\begin{figure}[t]
%%%%%%%%%%%%%%%%%%%%%%%%%%%%%%%%%%%%%%%%%%%%%%%%%%%%%%%%%%%%%%%%%%%%%%%%%%
%         \includegraphics[width=160mm]{dens_be8_4shl_rmcsm_yz01.eps}
  \includegraphics[width=\columnwidth]{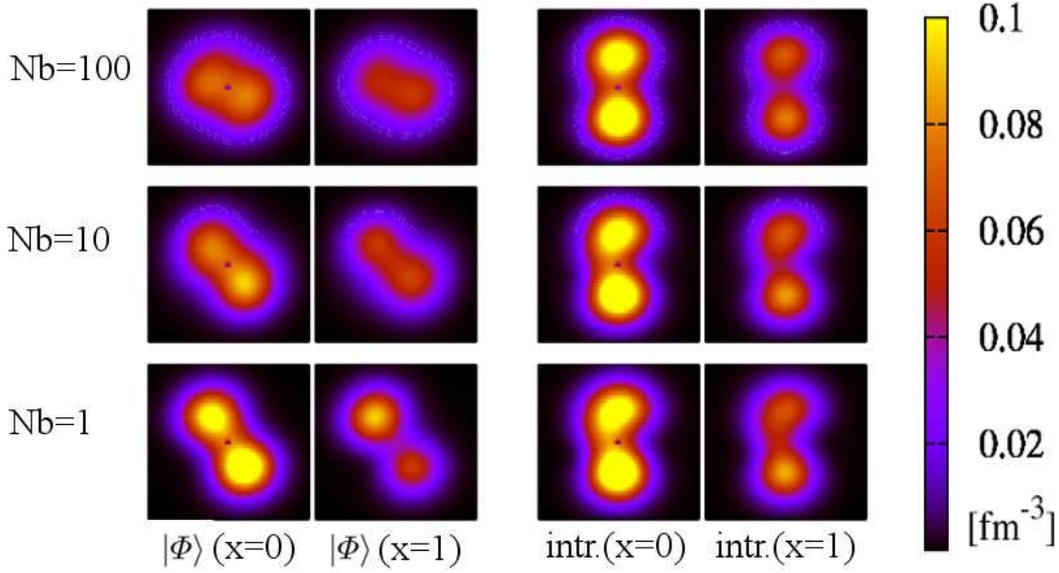}
%%%%%%%%%%%%%%%%%%%%%%%%%%%%%%%%%%%%%%%%%%%%%%%%%%%%%%%%%%%%%%%%%%%%%%%%%%
  \caption{$^8$Be proton density for $|\Phi\rangle$  % BR 
    (left panels) and intrinsic (right panels) states
    for various $N_b$ and sliced along the $yz$ plane.
    The number of basis states is $N_b=1, 10$ and $100$ for the lower, 
    middle and upper figures, respectively.
    The slice along the $yz$ plane is the $x=0$ fm plane (left) 
    or $x=1$ fm plane (right) for each panel.
    The size of each box is 8 fm $\times$ 8 fm.}
\label{DENSTAB}
\end{figure}

As an illustrative example, we show the $^{8}$Be density 
in $N_{shell}=4$ and $\hbar\omega=20$ MeV with the
JISP16 interaction for $J=0^+$ states.
The Coulomb interaction and the contamination of
spurious center-of-mass motion are neglected for simplicity. 
We show the proton density (a half of the total density) 
of the $|\Phi\rangle$ 
and the intrinsic-state density, $\rho^{{\rm intr}}$, in Fig. \ref{DENSTAB}.

The number of basis states is $N_b=1, 10$ and $100$ 
for the lower, middle and upper rows, respectively.
The energy is almost converged at $N_b=100$.
Each density distribution is shown 
along the $yz$ planes at $x=0$ fm and at $x=1$ fm.
%%%%%%%%%%%%%%%%%%%%%%%%%%%%%%%%%%%%%%%%%%%%%%%%%%%%%%%%%%%%%%%%%%%%%%%%%%
As shown in the $N_b=1$ results, clear deformation and the neck structure 
to be called a dumbbell shape appear. We can see that as the number of 
basis states, $N_b$, increases the density of $|\Phi\rangle$ are much vague and becomes 
ordinary prolate rather than dumbbell-like because of the mixture of 
different directions of principal axes of the basis states.
On the other hand, the intrinsic density has clearer dumbbell-like structure for each $N_b$.
In addition, the density distribution of the intrinsic state is almost unchanged 
with respect to $N_b$.
This result indicates the appearance of cluster structure in the no-core MCSM.
We also check how the cluster shape differs between $N_{shell}=3$ and $N_{shell}=4$.
We find that the neck of dumbbell shape is more enhanced in $N_{shell}=4$ 
than in $N_{shell}=3$.
% when the model space includes $pf$-orbit of $N_{shell}=4$.
%This is due to the effect of node at the center of the coordinate.
Since the weights of distribution for both sides of the principal axis are almost the same, this cluster can be considered as two $\alpha$ clusters.
The stability of the $\alpha$ cluster is confirmed with respect to $N_b$ and $N_{shell}$.
This result is consistent with the result of 
the Green's Function Monte Carlo \cite{yoshida:2000wp}. 
% The wave function of No-Core MCSM has advantage because it makes unified description for the ground state 
% and the excited states. %consistency to the ground state wave function. 
% This advantage is important when we analyzed excited states as a further application of this study. 
%%
%It is also important to see whether the appearance of $\alpha$ or any cluster state
% is stable with respect to $N_{shell}$ and $\hbar \omega$ as discussed in this subsection. 
With the use of this method to draw the density, 
we can study the appearance of cluster structure 
directly not only for $N=Z$ nuclei but also for the neutron-rich nuclei 
in the {\it ab initio} approach.
The study of exotic structure including unstable nuclei in the $p$-shell region is in progress.

%For further extension of this method, $J>0$ state should be investigated. The density can also be defined slightly complicated calculation than \ref{DENS} by taking $K$-mixing effect into account.

%%%%%%%%%%%%%%%%%%%%%%%%%%
% IF POSSIBLE BE ISOTOPES A=8 10 12 SHOULD BE ADDED
% IF POSSIBLE EFFECT OF CM MOTION SHOULD BE ADDED
% IF POSSIBLE PARITY INTRINSIC STATE SHOULD BE ADDED
%%%%%%%%%%%%%%%%%%%%%%%%%%%%%%%%%%%%%%%%%%%%%%%%%%%%%%%%%%%%%%%%%%%%%%%%%%
% MEMO
%However, the parity asymmetric distribution is not clearly seen in this definition,
%Therefore, we also use parity intrinsic density $\rho^{intr'}$, where another rotation $\beta=\pi$ is operated to $\Phi(D^R_n)$.
% ETC ETC 
%%%%%%%%%%%%%%%%%%%%%%%%%%%%%%%%%%%%%%%%%%%%%%%%%%%%%%%%%%%%%%%%%%%%%%%%%%

% \newpage

%%%% Sect.3.2 no-core calc. intrinsic state T. Yoshida
%%% % \input{sect32_yoshida.tex} % it is included in sect3_tabe.tex

%%%% Sect.4 Ni isotopes Y. Tsunoda
%\input{sect4_ytsunoda.tex}
\section{ Application to neutron-rich Cr and Ni isotopes}
\label{sect4_ytsunoda}
In this section, we discuss the application of the MCSM to 
the large-scale shell-model calculations about neutron-rich Cr and Ni isotopes 
as examples. We take a model space as the $pfg_9d_5$ shell, 
which consists of the $0f1p$ shell, the $0g_{9/2}$ orbit, and the $1d_{5/2}$ orbit.
By using such a sufficiently large model space, 
we aim at a unified description of medium-heavy nuclei 
and at studying the shell evolution 
\cite{otsuka_tensor_2001, otsuka_tensor_2005, 
  otsuka_tensor_2010, otsuka_threebody_2010}
and the magicity of  $N=28,40, 50$ 
microscopically.

\subsection{Ni isotopes and magicity of $N=28,40, 50$.  }

The nuclear shell structure evolves in neutron-rich nuclei 
and the magic numbers of unstable nuclei are different from those of stable nuclei.
The large excitation energy of the $2^+$ yrast state and 
the small $B(E2;0^+ \rightarrow 2^+)$ value 
in $^{68}$Ni ($Z=28$, $N=40$) might indicate that $^{68}$Ni is a double-magic nucleus, 
although $N=40$ is a magic number of the harmonic oscillator, not a magic number of 
the nuclear shell model. 
On the other hand, the small excitation energies of the $2^+$ yrast state 
and the large $B(E2;0^+ \rightarrow 2^+)$ values in Cr ($Z=24$) isotopes of $N\sim 40$ 
suggest rather strong deformation.
This change of the $N=40$ gap has been studied theoretically \cite{Nowacki}.
$^{78}$Ni, which has $Z=28$ and $N=50$ doubly magic numbers, 
has also been investigated 
to discuss its magicity  and the size of $N=50$ gap \cite{78Ni}.

In the $sd$-shell and the light $pf$-shell regions,
we can describe properties of stable nuclei in relatively small model spaces.
However, we sometimes need a large model space 
to describe the properties of neutron-rich nuclei. 
In order to discuss neutron-rich Ni isotopes up to $N=50$,
it is essential to include the effects of excitation 
across the $Z=28$ and $N=50$ gaps
by adopting the $pfg_9d_5$ model space.
Concerning this model space, 
M. Honma {\it et al.} proposed the A3DA effective interaction \cite{A3DA} 
which consists of the GXPF1A~\cite{GXPF1A}, JUN45~\cite{jun45}, 
and $G$-matrix effective interactions with phenomenological modifications.
It has succeeded in describing the neutron-rich Cr and Ni isotopes 
under a severe truncation of the model space utilizing 
the few-dimensional basis approximation  \cite{fda}.
In this work, we use the new version of the MCSM method, 
which enables us to precisely evaluate the exact shell-model energy 
without any truncation and discuss the effective interaction.

\subsection{ Effective interaction for $pfg_9d_5$ shell}

In this section, we discuss the  A3DA effective interaction~\cite{A3DA} 
and its improvement.
The two-body matrix elements (TBMEs) of the A3DA interaction consist of three parts.
The TBMEs of the $pf$ shell are those of the GXPF1A interaction~\cite{GXPF1A}, 
which is successful for describing spectroscopic properties of light $pf$-shell nuclei. 
The TBMEs of the $f_5pg_9$ shell related to the $0g_{9/2}$ orbit 
are those of the JUN45 interaction~\cite{jun45}.
The GXPF1A and JUN45 interactions were determined 
by combining microscopically derived interactions ($G$ matrix) with 
a minor empirical fit 
so as to reproduce experimental data.
The other TBMEs are from the $G$-matrix effective interaction\cite{Gmatrix,Hjorth-Jensen}, 
which is calculated 
from the chiral N3LO interaction\cite{N3LO}.
The Coulomb interaction is not included and the isospin symmetry is conserved.
The $G$ matrix is calculated for the $pfsdg$ shell with $^{40}$Ca as an inert core 
and the core-polarization correction is included perturbatively.
The single-particle energies and the monopole interaction are adjusted 
to reproduce the GXPF1A and JUN45 predictions for the $pf$ shell and $g_{9/2}$ orbits, 
and the Woods-Saxon single-particle energies 
of stable semi-magic nuclei for the other part. 

The original A3DA interaction failed to describe 
some nuclei around $N\sim 40$. 
We modify mainly single-particle energies
and monopole components related to the $0g_{9/2}$ orbit 
by comparing the results of the calculations with the experiments. 
These calculations are far beyond the current limitation 
of the conventional diagonalization method, and 
the MCSM method enables us to perform this comparison.

\subsection{ MCSM results of the neutron-rich Cr and Ni isotopes }

We performed systematic calculations of the $0^+$ and $2^+$ yrast states of
neutron-rich Cr and Ni even-even isotopes using the MCSM method and 
the modified A3DA interaction.
We took 50 basis states for the MCSM with the refinement procedure, 
which is discussed in Sect.\ref{sec:var}. 
The energies were extrapolated by the energy-variance extrapolation method 
and the other values were not.
The effective charges are taken as $(e_p, e_n) = (1.5, 0.5)e$.

\begin{figure}
  \rotatebox{-90}{
    \includegraphics[scale=0.4]{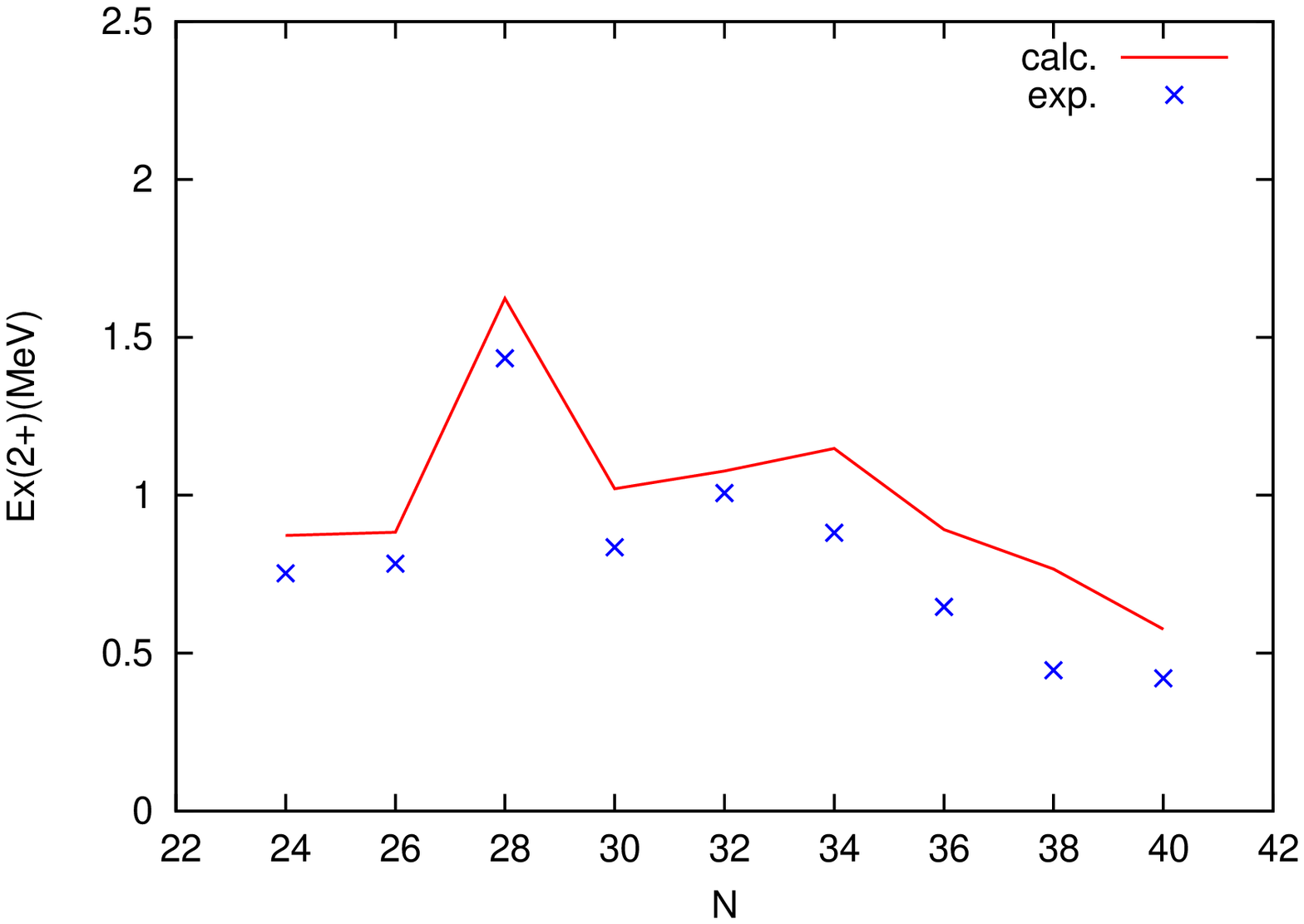}
  }
  \rotatebox{-90}{
    \includegraphics[scale=0.4]{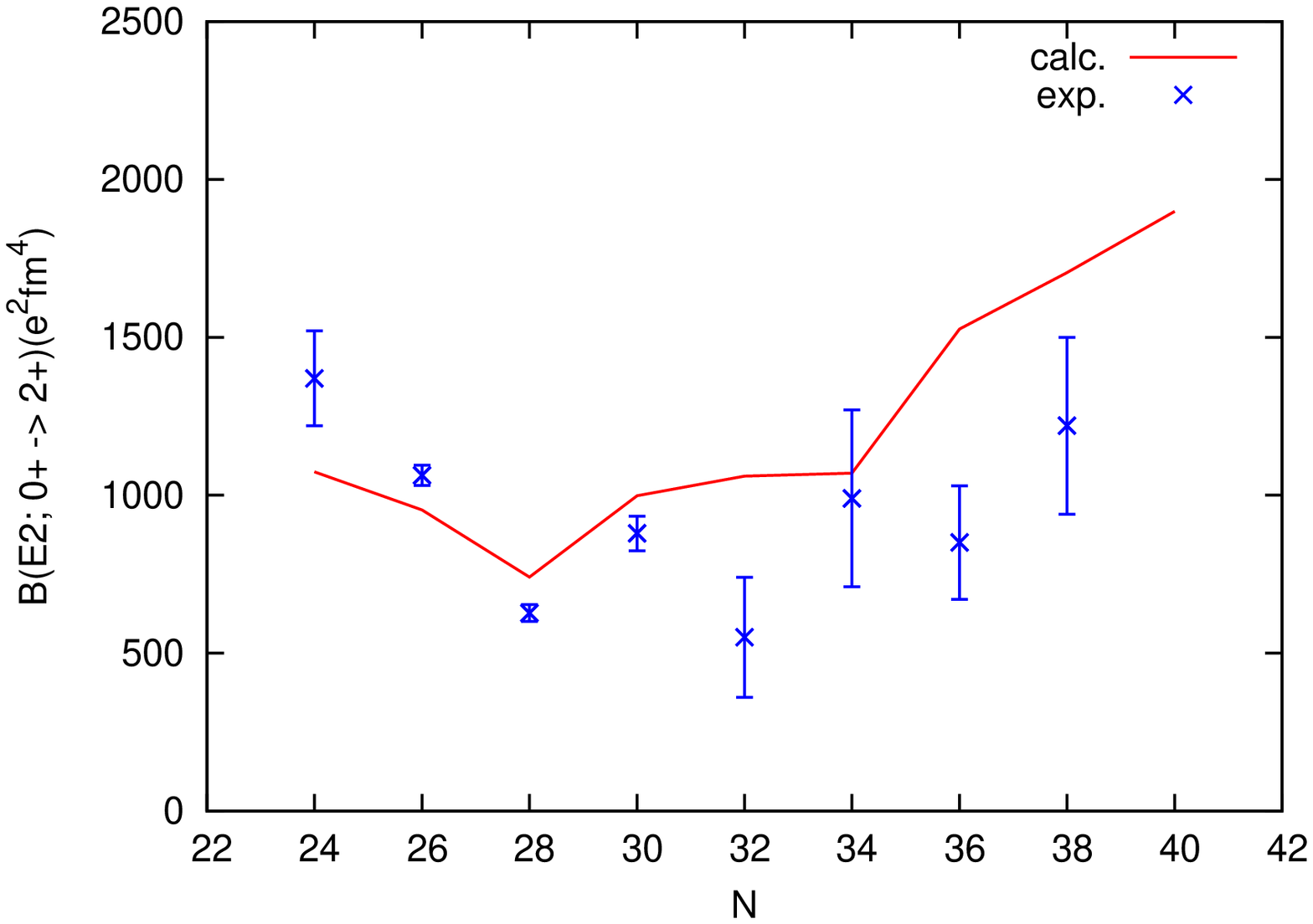}
  }
  \caption{The excitation energies of $2^+_1$ states (left) 
    and $B(E2;0^+_1 \rightarrow 2^+_1)$ values (right) obtained by the MCSM 
    for Cr isotopes. 
    Experimental data are taken from Refs.\cite{nudat, BE2}.}
  \label{fig:cr}
\end{figure}

Figure \ref{fig:cr} shows the $2^+$ excitation energies 
and the $B(E2)$ transition probabilities of neutron-rich Cr isotopes.
The MCSM results well reproduce the experimental values
while the modest overestimation remains.
The Cr isotopes do not show any feature of $N=40$ magicity, 
while the characteristics of $N=28$ magicity can be seen, 
namely, a sudden increase of excitation energy and slight decrease of 
the $B(E2)$ value. 
On the neutron-rich side, 
the excitation energy decreases and the $B(E2)$ value 
increases gradually as the neutron number increases, which implies
gradual enhancement of the quadrupole deformation.

\begin{figure}[tp]
 \begin{minipage}{0.48\hsize}
  \begin{center}
  \rotatebox{-90}{
   \includegraphics[height=\textwidth]{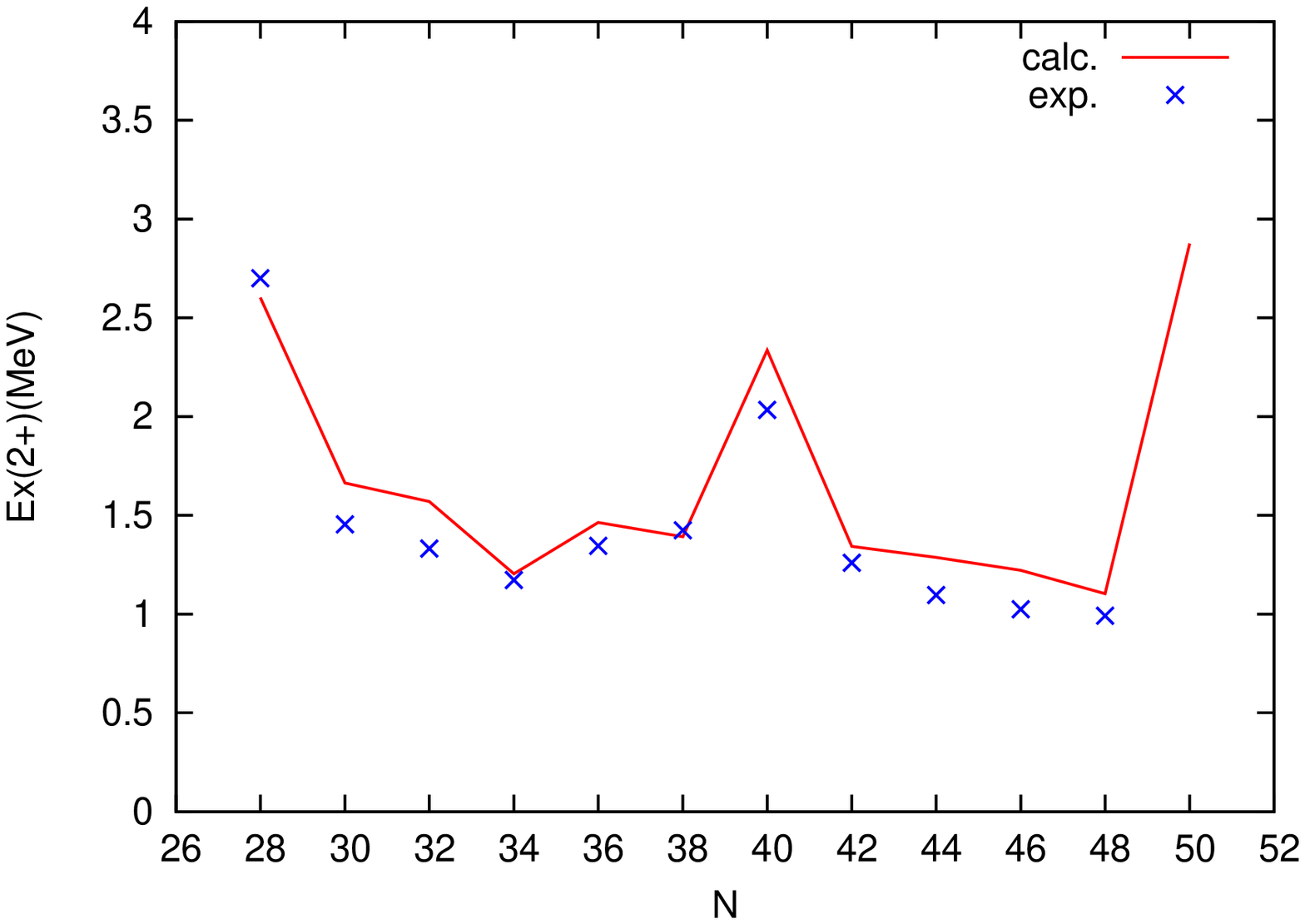}
  }
   \caption{$2^+$ excitation energies for Ni isotopes. 
     Experimental data are taken from Ref.\citen{nudat}.}
   \label{fig:Ni-ex}
  \end{center}
 \end{minipage}
 \hspace*{5mm}
 \begin{minipage}{0.48\hsize}
  \begin{center}
  \rotatebox{-90}{
   \includegraphics[height=\textwidth]{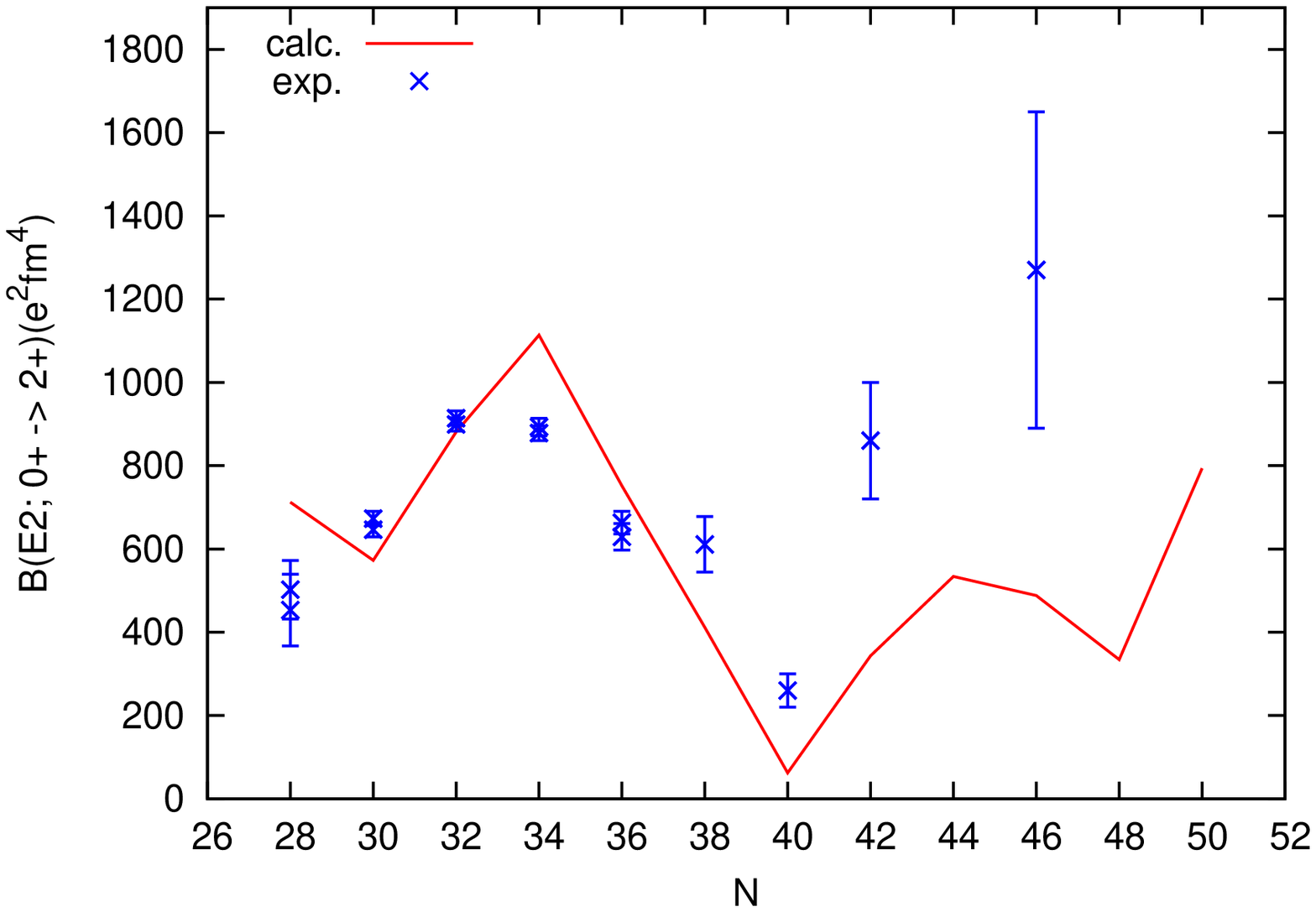}
  }
   \caption{$B(E2;0^+ \rightarrow 2^+)$ values for Ni isotopes.
            Experimental data are taken from Ref.\citen{BE2}.}
   \label{fig:Ni-BE2}
  \end{center}
 \end{minipage}
\end{figure}

\begin{figure}[tp]
 \begin{minipage}{0.48\hsize}
  \begin{center}
  \rotatebox{-90}{
   \includegraphics[height=\textwidth]{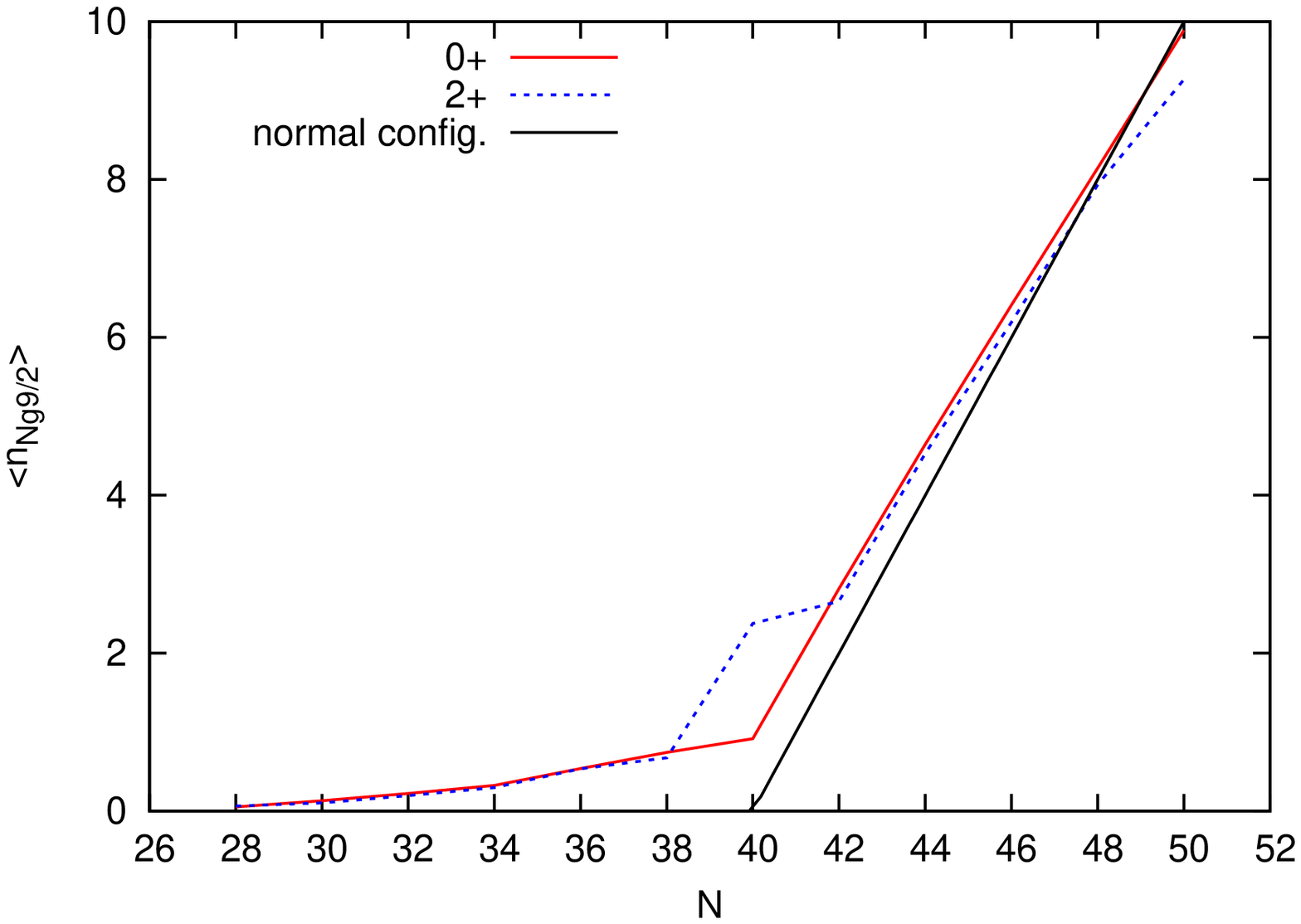}
  }
   \caption{Occupation numbers of the neutron $g_{9/2}$ orbit for Ni isotopes.}
   \label{fig:Ni-ng9}
  \end{center}
 \end{minipage}
 \hspace*{5mm}
 \begin{minipage}{0.48\hsize}
  \begin{center}
  \rotatebox{-90}{
   \includegraphics[height=\textwidth]{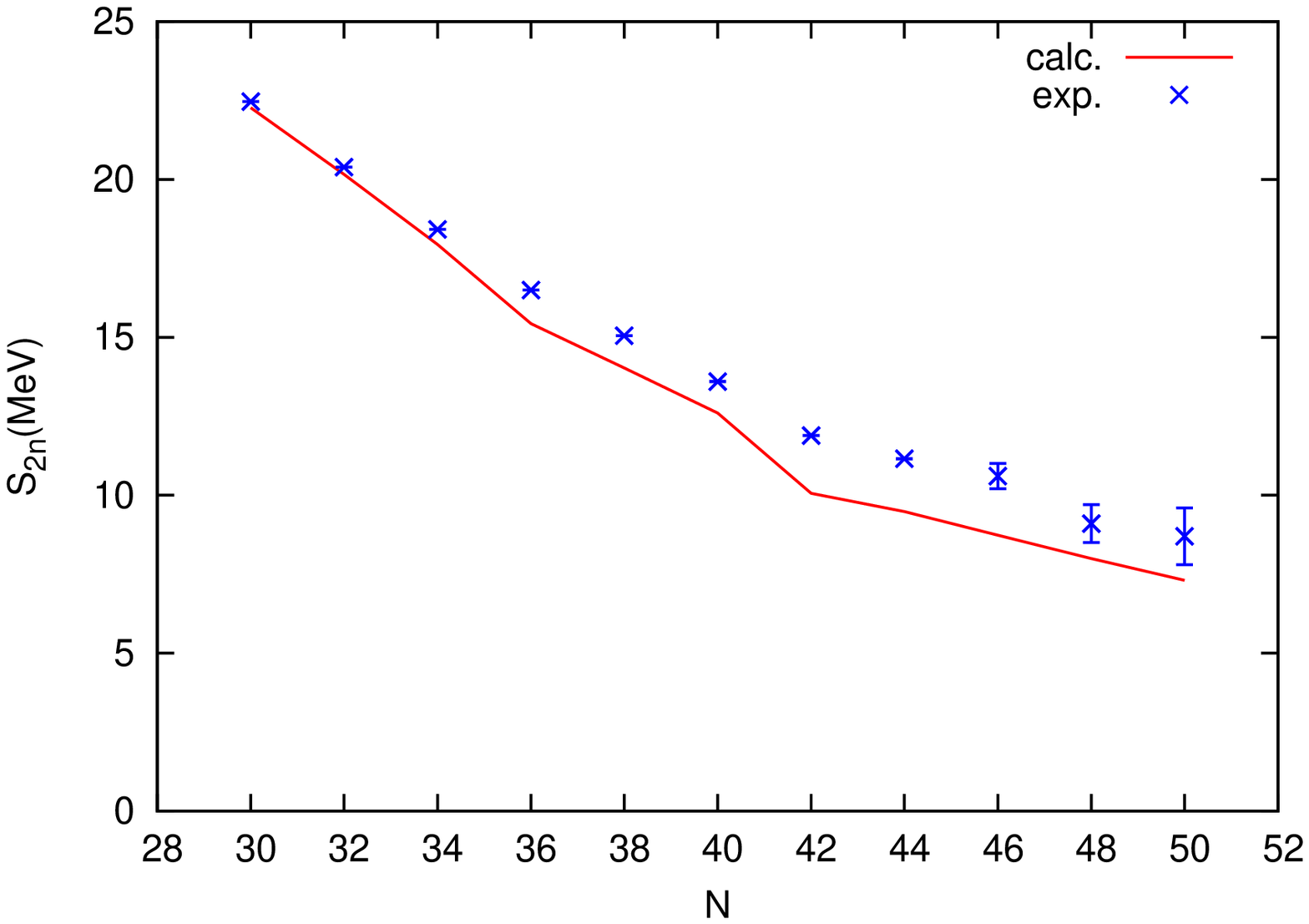}
  }
   \caption{Two-neutron separation energies $S_{2n}$ for Ni isotopes.
            Experimental data are taken from Ref.\citen{nudat}.}
   \label{fig:Ni-s2n}
  \end{center}
 \end{minipage}
\end{figure}

\begin{figure}[tp]
 \begin{center}
  \begin{minipage}{0.45\hsize}
   \includegraphics[width=\textwidth]{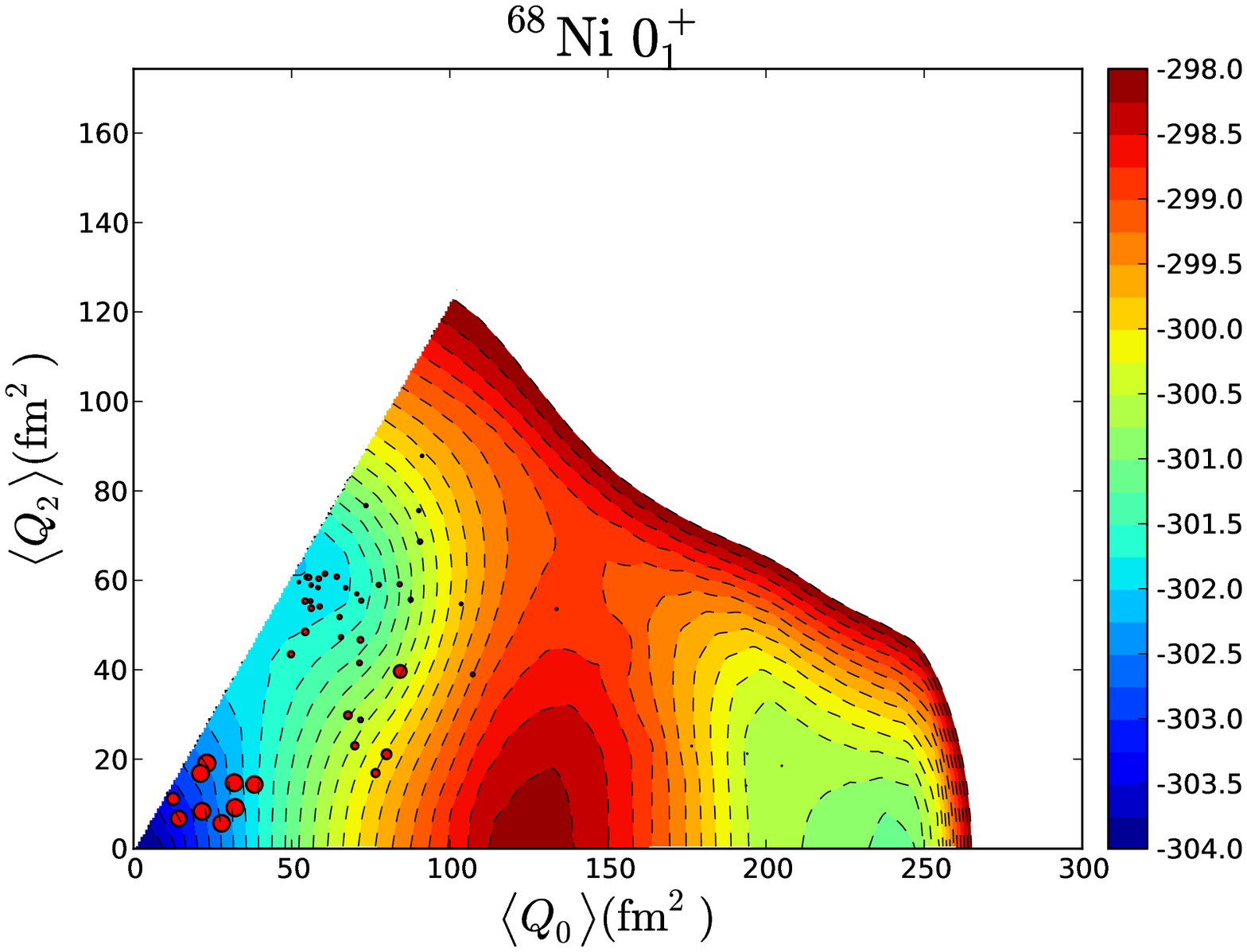}
  \end{minipage}
  \begin{minipage}{0.45\hsize}
   \includegraphics[width=\textwidth]{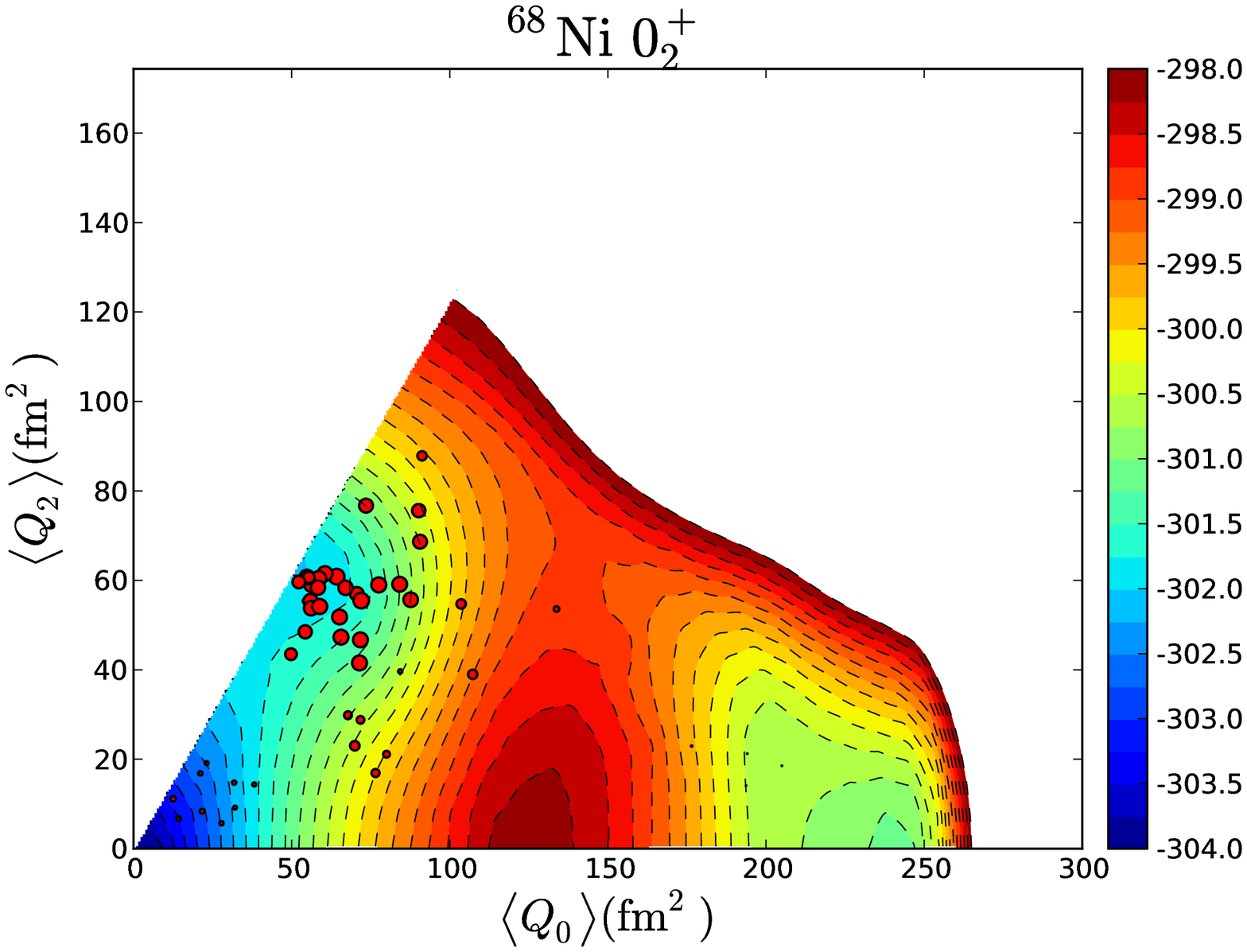}
  \end{minipage}
  \caption{Total energy surface of $0^+_1$ (left) and $0^+_2$ (right) states of  $^{68}$Ni. 
    The positions of red circles represent quadrupole deformations of the MCSM basis states
    before projection.
    The areas of those circles represent the overlap probabilities 
    of the basis states and the resulting wave function.}
  \label{fig:Ni-PES}
 \end{center}
\end{figure}

Figure \ref{fig:Ni-ex} shows $2^+$ excitation energies
of Ni ($Z=28$) even-even isotopes from $^{56}$Ni to $^{78}$Ni.
The large $2^+$ excitation energy of $^{56}$Ni ($N=28$) indicates $Z=28$, $N=28$ double magicity. 
The large value of the calculated $2^+$ excitation energy of $^{78}$Ni ($N=50$) suggests 
$Z=28$, $N=50$ double magicity. 
The large $2^+$ excitation energy of $^{68}$Ni ($N=40$) indicates $N=40$ magicity.
The calculated values reproduce the experimental values well.

Figure \ref{fig:Ni-BE2} shows $B(E2;0^+ \rightarrow 2^+)$ for neutron-rich Ni isotopes.
The small value of $B(E2;0^+ \rightarrow 2^+)$ at $N=40$ indicates $N=40$ magicity. 
Neither the theoretical nor experimental value of $B(E2;0^+ \rightarrow 2^+)$ 
at $N=28$ is small unlike that at the $N=40$, 
and the theoretical $B(E2;0^+ \rightarrow 2^+)$ value at $N=50$ becomes large 
in comparison with those of neighboring nuclei.
It suggests that at $N=28,50$ magicity is broken to some extent for $^{56,78}$Ni, respectively.
Figure \ref{fig:Ni-ng9} shows the occupation number of the neutron $g_{9/2}$ orbit. 
The occupation numbers of $0^+$ and $2^+$ states are very close for Ni isotopes besides $^{68,78}$Ni ($N=40,50$).
The occupation numbers of the $2^+$ states of $^{68,78}$Ni 
show a breakdown of the closed-shell structure.
Figure \ref{fig:Ni-s2n} shows two-neutron separation energies.
The calculated values of neutron-rich nuclei are smaller than experimental values.
This means that the binding energies of neutron-rich nuclei are underestimated.
The values of $S_{2n}$ increase slightly by considering the Coulomb energy, 
but calculated values are still smaller than experimental values.

Figure \ref{fig:Ni-PES} shows the total energy surface of $^{68}$Ni 
provided by the $Q$-constrained Hartree-Fock calculation \cite{ringschuck}.
There are three minimum points for $^{68}$Ni.
Figure \ref{fig:Ni-PES} also shows quadrupole deformations of the 
MCSM wave functions of the $0^+_{1}$ and $0^+_{2}$ states.
The scattered circles correspond to the basis states in the MCSM wave function.
The position of the circle indicates the quadrupole deformation of the basis state before projection.
The area of the circle is proportional to the overlap probability of the projected basis and the resulting wave function.
It is quite clear that the $0^+_1$ state of $^{68}$Ni corresponds to a spherical shape and the $0^+_2$ state corresponds to an oblate shape.
Spherical and oblate components are mixed to some extent, 
but the components of the prolate minimum hardly mix.

\begin{figure}[tp]
 \begin{minipage}{0.48\hsize}
  \begin{center}
  \rotatebox{-90}{
   \includegraphics[height=\textwidth]{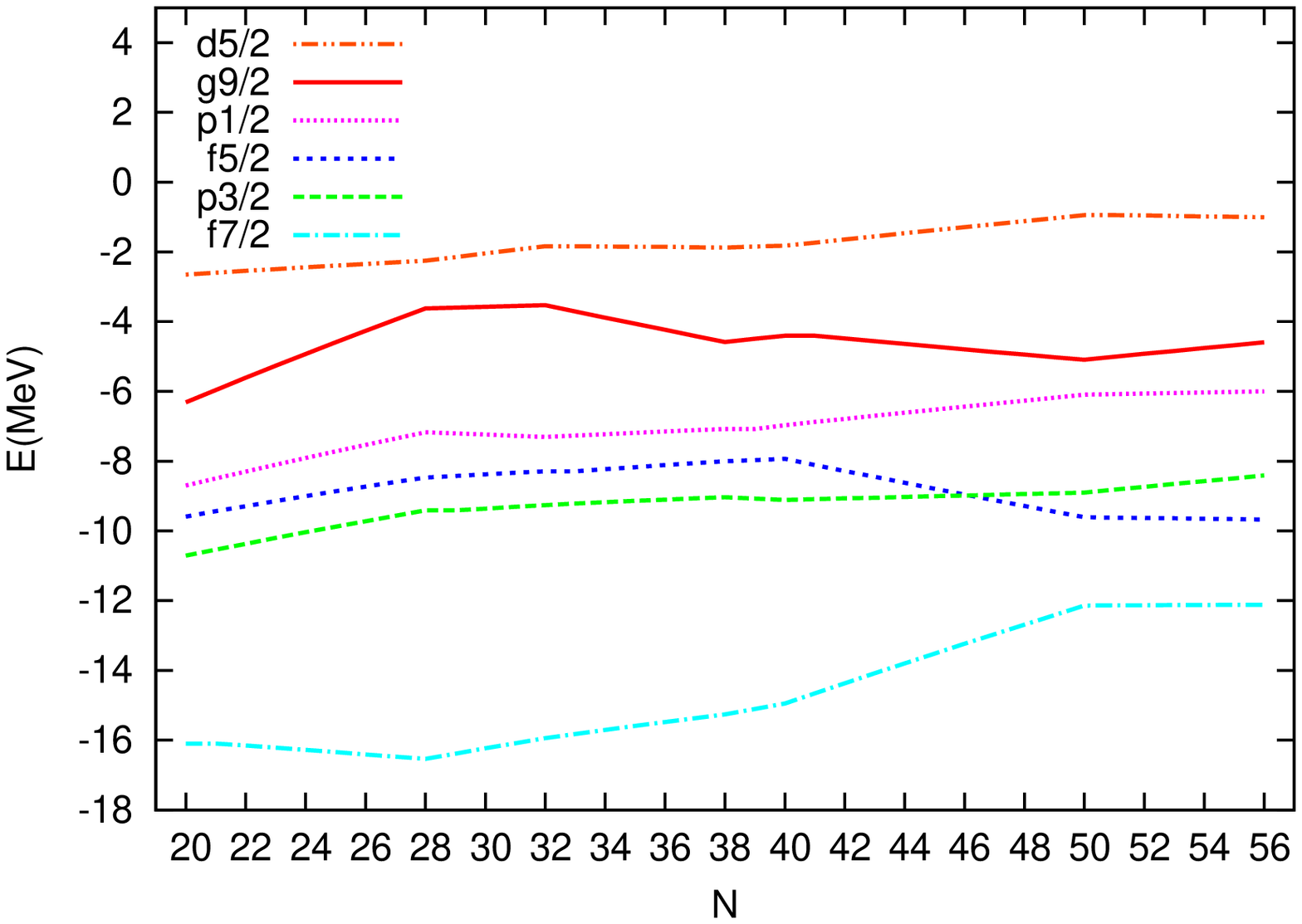}
  }
   \caption{ESPEs of the neutron orbits for Ni ($Z=28$) isotopes.}
   \label{fig:ESPE-Ni-n}
  \end{center}
 \end{minipage}
 \hspace*{5mm}
 \begin{minipage}{0.48\hsize}
  \begin{center}
  \rotatebox{-90}{
   \includegraphics[height=\textwidth]{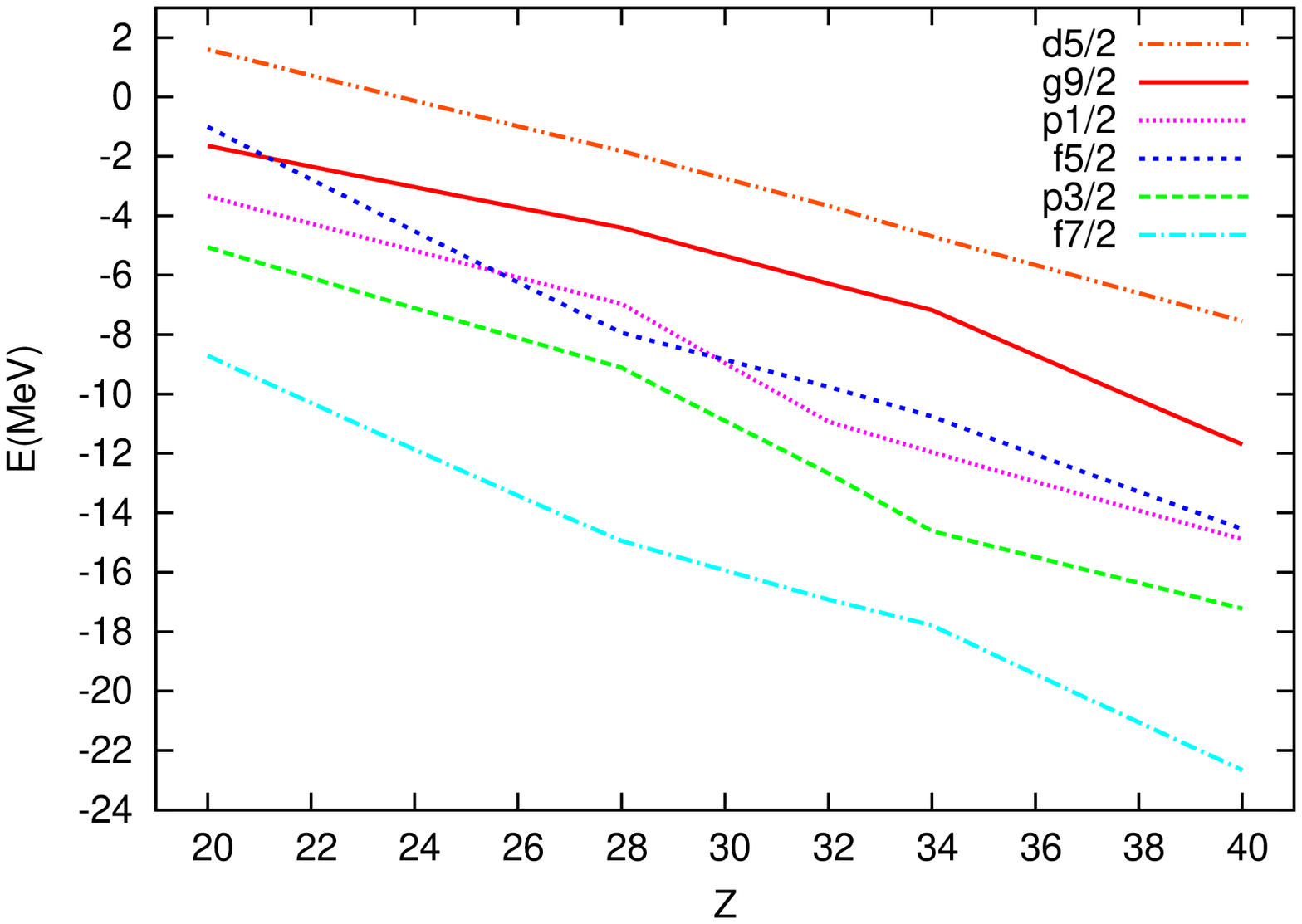}
  }
   \caption{ESPEs of the neutron orbits for $N=40$ isotones.}
   \label{fig:ESPE-N40-n}
  \end{center}
 \end{minipage}
\end{figure}

Furthermore, we consider the magicity and the energy gaps for Ni isotopes 
by using the effective single particle energies (ESPEs) \cite{utsuno_espe}. 
Figure \ref{fig:ESPE-Ni-n} shows ESPEs of the neutron orbits for Ni isotopes. 
The $f_{7/2}$-$p_{3/2}$ gap at $N=28$ is $7.1$ MeV and gives the magicity to $^{56}$Ni.
The $g_{9/2}$-$d_{5/2}$ gap at $N=50$ is $4.2$ MeV  and gives the magicity to $^{78}$Ni.
This is partly due to the additional lowering of the $g_{9/2}$ orbit
caused by pairing correlation between 
two neutrons in the $g_{9/2}$ orbit, and also due to the effect of the two-neutron 
repulsive monopole interaction originating in the three-nucleon 
force like in exotic oxygen isotopes \cite{otsuka_threebody_2010}.
The $p_{1/2}$-$g_{9/2}$ gap at $N=40$ is $2.6$ MeV, which is smaller than the $N=28,50$ gaps.
Figure \ref{fig:ESPE-N40-n} shows the ESPEs of the neutron orbits for $N=40$ isotones. 
As the proton number of $f_{7/2}$ increases from $Z=20$ to $Z=28$, 
the ESPE of $f_{5/2}$ lowers and the $N=40$ gap becomes larger.
Because of this evolution of the $N=40$ gap, 
the properties of $N\sim 40$ nuclei depend on the proton number.

\begin{figure}
  \rotatebox{-90}{
    \includegraphics[scale=0.4]{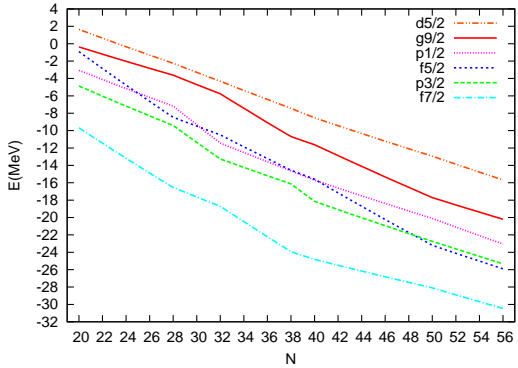}
  }
  \rotatebox{-90}{
    \includegraphics[scale=0.4]{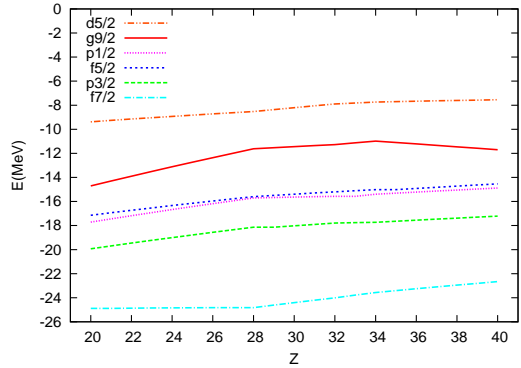}
  }
  \caption{
    ESPEs of the proton orbits for Ni isotopes (left) 
    and for $N=40$ isotones (right).
  }
  \label{fig:proton_espe}
\end{figure}

In Fig. \ref{fig:proton_espe}, 
the ESPEs of the proton orbits for Ni isotopes 
and for $N=40$ isotones are shown. In the former, rapid lowering 
of the $f_{5/2}$ orbit from $N=40$ to 50 is clearly seen as suggested in 
\cite{otsuka_tensor_2005, otsuka_tensor_2010},
while narrowing of $Z=28$ gap is also visible there.
Such changes are responsible partly for the origins of the structure evolution 
in these Ni isotopes.

%%%% Sect.5 Summary, Appendix  Shimizu
%\input{sect5_summary.tex}

\section{Summary and future perspectives}
\label{sect5_summary}

We have developed a new generation of the MCSM by introducing 
the conjugate gradient method and the energy-variance extrapolation, 
which enhance the applicability of the MCSM greatly.
We have two major scopes of this framework: {\it ab initio} shell-model 
calculations and conventional shell-model calculations assuming an inert core.
In the former, we have compared the MCSM results with the exact FCI
calculations to demonstrate the validity of the MCSM framework 
and its feasibility beyond the limit of the FCI in Sect.\ref{sect3_tabe}. 
In addition, we have proposed a novel method to discuss the intrinsic 
structure and demonstrated that the cluster structure appears 
in shell-model-type calculations based on the harmonic-oscillator-basis 
wave function.
In the latter, we discussed in Sect.\ref{sect4_ytsunoda}
that the MCSM enables us to perform shell-model calculations
of neutron-rich Cr and Ni isotopes 
in the $pfg_9d_5$ model space in which the isospin symmetry is conserved.
We proposed a ``modified A3DA'' interaction which 
reproduces the low-lying spectra of neutron-rich Cr and Ni isotopes 
and guides us towards a unified description including  $^{56}$Ni, $^{68}$Ni
and $^{78}$Ni, with magic numbers 28, 40, 50, respectively.
The prediction of $^{78}$Ni is especially 
interesting to see the evolution of shell structure. 
On the other hand,  Cr isotopes do not show any feature of $N=40$ magicity 
and the collectivity enhances as the neutron number increases.
The  MCSM and newly proposed effective interaction 
are expected to provide us with a unified description of $pf$-shell nuclei.

The current status of the computer-code development was also 
reported in Sect.\ref{sec:computational_aspects}. 
At the present stage,
we have obtained good parallel scalability of our code 
up to $10^5$ CPU cores 
via early access to the K computer at RIKEN AICS\cite{kcomputer} 
as measured by the benchmark test.
However, such good scalability is not always obtained 
and further development is in progress. 
This activity is promoted strongly 
as a part of the activities of 
HPCI Strategic Programs for Innovative Research (SPIRE) Field 5 
``The origin of matter and the universe''.

By utilizing both the developed code and the K computer, 
we promote further large-scale shell-model calculations
as a part of the SPIRE activities.
We plan to perform systematic study with {\it ab initio} calculations
of light nuclei in $N_{shell}=5$ and some states in  $N_{shell} =6$.
Concerning the medium-heavy nuclei, because it is difficult to 
cover whole region of the nuclear chart, we will choose some 
interesting nuclides
as subjects of our investigation, and will 
perform shell-model calculations of these nuclides
with the two-major-shell model space. 
For example, the shell-model calculations of 
$^{130}$Te, $^{128}$Te,  and $^{150}$Nd 
are extremely interesting to study double beta decay and 
the nuclear matrix element of neutrinoless decay.
We also continue to study the systematic calculations of 
neutron-rich $pf$-shell nuclei
to discuss the shell-evolution phenomenon.

\section*{Acknowledgments}
We acknowledge Professors J. P. Vary, P. Maris and Dr. L. Liu for 
our collaboration concerning {\it ab initio} shell-model calculations. 
This work has been supported by Grants-in-Aids
for Scientific Research  (23244049), 
for Scientific Research on Innovative Areas (20105003), 
and for Young Scientists (20740127) from JSPS,
the SPIRE Field 5 from MEXT, and the CNS-RIKEN joint project
for large-scale nuclear structure calculations.
The numerical calculations were performed mainly
on the T2K Open Supercomputers at the University of Tokyo
and Tsukuba University.
The exact conventional shell-model calculations were performed
by the code MSHELL64 \cite{mshell64}.

\appendix

\section{Numeration with projected Slater determinants}
\label{sec:eqs}
In this appendix, we show some equations 
which are needed to perform the calculation discussed in Sect.\ref{sect2_shimizu}.

At the beginning, we define a deformed Slater determinant, 
\begin{equation}
  |\phi \rangle = \prod_{k=1}^{N_{\rm f}} 
  \left( \sum_{l=1}^{N_{\rm sp}} D_{lk} c^\dagger_l \right) | - \rangle , 
  \label{eq:slater}
\end{equation}
which is parametrized by the complex $N_{\rm sp} \times N_{\rm f}$ matrix $D$ with 
the normalization condition $D^\dagger D=1$.
$N_{\rm f}$ and $N_{\rm sp}$ are the numbers of fermions and single-particle states, 
respectively. 
The $|-\rangle$ denotes an inert core 
in the conventional shell-model calculations 
or the vacuum in {\it ab initio} shell-model calculations.
Because  we do not mix the proton and neutron space in practical calculations, 
the wave function is written as a product of proton and neutron Slater determinants, 
namely, 
$|\phi\rangle = |\phi_{\rm proton}\rangle \otimes |\phi_{\rm neutron} \rangle$.
For simplicity,  we do not write this isospin degree of freedom explicitly.
One can easily reproduce the equations representing 
the explicit proton-neutron degree of 
freedom by taking $D$ of the proton-neutron sector as zero such as
\begin{equation}
  \label{eq:d-pn}
   D = \left(
     \begin{array}{cc}
       D^\pi & 0 \\
       0  & D^\nu
     \end{array}
     \right)
\end{equation}
where $D^\pi$ and $D^\nu$ represent Slater determinants of protons and neutrons, 
respectively.

The angular-momentum, parity projector $ P^{I\pi}_{MK} $ 
in Eq.(\ref{eq:proj}) is 
performed by discretizing the integral concerning 
the Euler angles such as
\begin{equation}
  \label{eq:proj_sum}
  P^{I\pi}_{MK} = \sum_\lambda W^{I\pi(\lambda)}_{MK} R^{(\lambda)}
\end{equation}
where the $\lambda$ denotes an index of mesh point 
of the discretization (here, a set of 
Euler's angle $\Omega=(\alpha, \beta, \gamma)$ 
and parity variable $\pi^{\lambda} = \pm 1$). 
In this paper, the parity projection is described 
by the summation of  2 mesh points such as 
$P^\pi = \frac{1 + \pi \Pi}{2} = \sum_{\lambda=1}^2 \pi^{(\lambda)}\Pi^{(\lambda)}$
with $\pi^{(1)}=\frac{1}{2}$, $\pi^{(2)}=\frac{\pi}{2}$, 
$\Pi^{(1)}=1$, and $\Pi^{(2)}=\Pi$ with $\Pi$ being the parity-conversion operator.
$W^{(\lambda)}$ is a weight of the mesh point $\lambda$, 
and $R^{(\lambda)}$ is a product of 
the rotation and parity-conversion operators such as
\begin{eqnarray}
  W^{I\pi(\lambda)}_{MK} &=& 
  \frac{2I+1}{8\pi^2} D^{I*}_{MK}(\alpha_\lambda, \beta_\lambda, \gamma_\lambda)
  \pi^{(\lambda)} ,  \nonumber \\
  R^{(\lambda)} &=& e^{i\alpha_\lambda J_z} e^{i\beta_\lambda J_y} e^{i\gamma_\lambda J_z} 
  \Pi^{(\lambda)}. 
\end{eqnarray}
Note that the operator $R^{\lambda}$ 
does not change the form of a Slater determinant, i.e., 
\begin{equation}
  |\phi^{(\lambda)}_n \rangle = R^{(\lambda)} | \phi_n \rangle , 
\end{equation}
where a matrix $D^{n(\lambda)}$ represents the single Slater determinant 
$|\phi^{(\lambda)}_n \rangle $, 
thanks to the Baker-Hausdorff's theorem \cite{ppnp_mcsm}.

The norm matrix and hamiltonian matrix spanned by $N$ Slater determinants
are written as 
\begin{eqnarray}
  \label{eq:norm-matrix}
  {\cal N}_{mM,nK} &=& \langle \phi_m| P^{I\pi}_{MK} | \phi_n \rangle  \\
  \label{eq:hamil-matrix}
  {\cal H}_{mM,nK} &=& \langle \phi_m| H P^{I\pi}_{MK} | \phi_n \rangle  .
\end{eqnarray}
The coefficient, $f_{nK}$ in Eq.(\ref{eq:wf}), is determined by 
solving the generalized eigenvalue problem
\begin{equation}
  \sum_{nK} {\cal H}_{mM,nK}f_{nK} 
  = {\cal E} \sum_{nK} {\cal N}_{mM,nK} f_{nK}, 
\end{equation}
and the normalization condition $\langle \Psi|\Psi\rangle = 1$.
The lowest eigenvalue of ${\cal E}$ is taken as $E_N$ 
if you would like to obtain the yrast state.

By combining Eqs. (\ref{eq:proj_sum}) and (\ref{eq:norm-matrix}),  
the norm matrix is calculated as 
\begin{equation}
  \label{eq:norm-mat-sum}
  {\cal N}_{mM,nK}  
  = \sum_\lambda W^{I\pi(\lambda)}_{MK} 
  \langle \phi_m | R^{(\lambda)} | \phi_n \rangle 
  = \sum_\lambda W^{I\pi(\lambda)}_{MK} 
  \langle \phi_m | \phi_n^{(\lambda)} \rangle , 
\end{equation}
with
\begin{equation}
  \label{eq:olp}
  \langle \phi_m | \phi_n^{(\lambda)} \rangle 
  =  {\rm det}\left( D^{m\dagger} D^{n(\lambda)} \right) .
\end{equation}

In the same way, the hamiltonian matrix is obtained as
\begin{eqnarray}
  \label{eq:ham-mat-sum}
  {\cal H}_{mM,nK}  &=& \sum_\lambda W^{I\pi(\lambda)}_{MK} 
  \langle \phi_m | H R^{(\lambda)} | \phi_n \rangle  \\
  &=& \sum_\lambda W^{I\pi(\lambda)}_{MK} 
  \langle \phi_m| \phi_n^{(\lambda)} \rangle
  {\rm Tr} \left(\rho^{(\lambda)} (t + \frac12 \Gamma^{(\lambda)}) \right) .
  \nonumber
\end{eqnarray}
where the generalized density matrix, $\rho^{(\lambda)}$, 
and the self-consistent field,  $\Gamma^{(\lambda)}$, \cite{ringschuck} are  defined as
\begin{equation}
  \label{eq:density_matrix}
  \rho^{(\lambda)}_{ij} = \frac{  \langle \phi_m| c^\dagger_j c_i| \phi^{(\lambda)}_n \rangle  }
  {  \langle \phi_m| \phi_n^{(\lambda)} \rangle  }
  = ( D^{n(\lambda)}(D^{m\dagger} D^{n(\lambda)})^{-1} D^{m\dagger} )_{ij}
\end{equation}
\begin{eqnarray}
  \Gamma^{(\lambda)}_{ik} &=&  \sum_{jl} \overline{v}_{ijkl}  \rho^{(\lambda)}_{lj} 
\end{eqnarray}
with $\overline{v}_{ijkl} = v_{ijkl}-v_{ijlk}$. 
The trivial summations and their indices 
for the matrix products are omitted for readability.
The indices $m,n$ of $\rho^{(\lambda)}$ and $\Gamma^{(\lambda)}$ are also omitted.

The most-time-consuming part is the calculation of the $\Gamma^{(\lambda)}_{ik}$, 
which can be rewritten following the idea of Ref.\cite{mcsm_tuning}, 
\begin{eqnarray}
  \label{eq:tb_matvec}
  \Gamma^{(\lambda)}_{a} &=&  \sum_{b} \overline{v}_{ab}  \rho^{(\lambda)}_{b} 
\end{eqnarray}
where  $a=(i,k)$, and $b=(j,l)$.
Because $\overline{v}_{ab}$ is a block-antidiagonal form 
owing to the symmetry of the Hamiltonian, 
Eq.(\ref{eq:tb_matvec}) is calculated as the products of 
the dense block matrices and the dense matrices in terms of 
the indices $a,b,\lambda$  efficiently 
avoiding trivial zero matrix elements of $v_{ab}$.
This method is referred to as the {\it matrix-matrix method} 
in Ref.\citen{mcsm_tuning}.
This {\it matrix-matrix method} enables us to use a CPU
utilizing the BLAS level 3 library quite efficiently, 
and the performance reaches  $70\sim 80\%$ of 
the theoretical peak performance \cite{mcsm_tuning}.

This efficient computation of $\Gamma^{(\lambda)}$ is useful also for 
the evaluation of the energy gradient, 
which is essential for the conjugate gradient method.
The energy gradient of the Slater-determinant coefficients is written as
\begin{eqnarray}
  \frac{\partial E_N}{\partial D^{m*}} 
  &=& (1-D^mD^{m\dagger}) \sum_{M,n,K,\lambda} f_{mM}^* f_{nK} 
  W^{I\pi(\lambda)}_{MK} \langle 
  \phi_m |\phi^{(\lambda)}_n \rangle 
  \\
  && \times
  \left(
    ( 1 - \rho^{(\lambda)})
    ( t + \Gamma^{(\lambda)} ) 
    + 
    \left( {\rm Tr}\bigl((t + 
      \frac12 \Gamma^{(\lambda)} \bigr)
      \rho^{(\lambda)}  \bigr) - E_N 
    \right)
  \right)  \rho^{(\lambda)} D^m.
  \nonumber 
\end{eqnarray}

To evaluate the energy variance 
$\langle \Delta H^2\rangle_N = \langle H^2\rangle_N  - E_N^2 $, 
the expectation value of the  $H^2$ 
with the wave function in Eq.(\ref{eq:wf}) is written as
\begin{eqnarray}
  \langle \Psi_N | H^2 |\Psi_N \rangle 
  &=& \sum_{m,M,n,K,\lambda}  f_{mM}^* f_{nK} W^{I\pi(\lambda)}_{MK} 
  \langle \phi_m| H^2 | \phi_n^{(\lambda)}\rangle .
  \label{eq:ex_val}
\end{eqnarray}
From Ref.\citen{mcsm_extrap}, 
the matrix element of the Hamiltonian squared is computed such as
\begin{eqnarray}
  \frac{ \langle \phi_m | H^2 | \phi^{(\lambda)}_n \rangle }
  { \langle \phi_m | \phi^{(\lambda)}_n  \rangle } 
  &=& 
  \sum_{i<j, k<l, \alpha<\beta, \gamma<\delta}
  v_{ijkl} \Theta^{(\lambda)}_{kl\alpha\beta} v_{\alpha\beta\gamma\delta} 
  \Lambda^{(\lambda)}_{\gamma\delta ij}
  \label{eq:variance} \\
  && + {\rm Tr}((t+\Gamma^{(\lambda)}) (1-\rho^{(\lambda)})(t+\Gamma^{(\lambda)}) \rho^{(\lambda)})
  + \left( {\rm Tr}(\rho^{(\lambda)}(t+\frac12 \Gamma^{(\lambda)}))\right)^2 
  \nonumber 
\end{eqnarray}
\begin{eqnarray}
  \Lambda^{(\lambda)}_{ijkl} 
  &=&  \rho^{(\lambda)}_{i k} \rho^{(\lambda)}_{j l} 
  - \rho^{(\lambda)}_{i l} \rho^{(\lambda)}_{j k}
\end{eqnarray}
\begin{eqnarray}
  \Theta^{(\lambda)}_{ijkl} 
  &=& (1- \rho^{(\lambda)})_{i k}(1-\rho^{(\lambda)})_{j l}  
  - (1- \rho^{(\lambda)})_{i l}(1-\rho^{(\lambda)})_{j k} .
\end{eqnarray}
The most-time-consuming part in the evaluation of the energy variance is 
to calculate the first term of the right-hand side of  Eq.(\ref{eq:variance}).
By substituting $(i,j), (k,l), (\alpha,\beta)$, 
and $(\gamma, \delta)$ by $a, b, c$, and $d$, respectively, 
this term is efficiently calculated as the products of the matrices, 
namely, $\sum_{abcd} v_{ab} \Theta_{bc} v_{cd} \Lambda_{da}$.
Note that the $v_{ab}$ has a block-diagonal form, which again enables 
us to use the BLAS level 3 library, avoiding trivial zero matrix elements.

Another formulation to compute 
the expectation values in projected Slater determinants
can be found in Refs.\cite{puddu_eqs, puddu_eff_variance},
in which the two-body interaction is decomposed into 
a sum of the squares of the one-body operators.

%%%% References
%\input{refs.tex}


\begin{thebibliography}{99}

%------------------------------------------------------------------------------
% References in Section 1 by Y. Utsuno
%------------------------------------------------------------------------------


\bibitem{kamada01}
H.~Kamada {\it et al.}, \PRC{64,2001,044001}. 


\bibitem{mj1949}
M.~G.~Mayer, \PR{75,1949,1969}; O.~Hazel, J.~H.~D.~Jensen, and 
H.~E.~Suess, \PR{75,1949,1766}. 

\bibitem{usd}
B.~A.~Brown and B.~H.~Wildenthal, Ann. Rev. Nucl. Part. Sci. {\bf 38}
	(1988), 29. 

\bibitem{usdab}
B.~A.~Brown and W.~A.~Richter, \PRC{74,2006,034315}. 

\bibitem{kb3g}
A.~Poves,  J.~S\'anchez-Solano, E.~Caurier, and F.~Nowacki, 
\NPA{694,2001,157}. 

\bibitem{gxpf1}
M.~Honma, T.~Otsuka, B.~A.~Brown, and T.~Mizusaki,
	\PRC{65,2002,061301(R)}; \PRC{69,2004,034335}. 

% Primary references for the NCSM - theory - including reactions
\bibitem{NCSM} 
P.~Navr{\'a}til, J.~P.~Vary, and B.~R.~Barrett, Phys.~Rev.~Lett.~{\bf 84} (2000), 5728;
Phys.~Rev.~C {\bf 62} (2000), 054311; 
S. Quaglioni and P.~Navr{\'a}til, Phys.~Rev.~Lett.~{\bf 101} (2008), 092501; 
Phys.~Rev.~C {\bf 79} (2009), 044606.

\bibitem{scidac}
SciDAC Review, Issue 6  (2007), pp. 42-51. 

% \bibitem{dimension} J.P.Vary, private communication.

\bibitem{mcsm_1996}
M.~Honma, T.~Mizusaki, and T.~Otsuka, \PRL{77,1996,3315}. 

\bibitem{mcsm_1995} M. Honma, T. Mizusaki, and T. Otsuka,
  Phys. Rev. Lett. {\bf 75} (1995), 1284.

\bibitem{smmc} S.~E.~Koonin, D.~J.~Dean, and K.~Langanke, 
Phys. Rep. {\bf 278} (1997), 1.

\bibitem{ppnp_mcsm}  T. Otsuka, M. Honma, T. Mizusaki, N. Shimizu,
  and Y. Utsuno, Prog. Part. Nucl. Phys. {\bf 47}  (2001), 319.


\bibitem{mcsm_pair}
N.~Shimizu, T.~Otsuka, T.~Mizusaki, and M.~Honma, 
\PRL{86,2001,1171}. 

\bibitem{vampier_2004}
K.~W.~Schmid, Prog. Part. Nucl. Phys. {\bf 52} (2004), 565. 

\bibitem{mcsm_1998}
T.~Otsuka, M.~Honma, and T.~Mizusaki, \PRL{81,1998,1588}. 

\bibitem{riken_rmcsm} N. Shimizu, Y. Utsuno, T. Abe, and T. Otsuka,
  RIKEN Accel. Prog. Rep. {\bf 43} (2010), 46.

\bibitem{mcsm_extrap} N. Shimizu, Y. Utsuno, T. Mizusaki, T. Otsuka, T. Abe,
  and M. Honma, Phys. Rev. C \textbf{82} (2010), 061305(R).

\bibitem{otsuka_tensor_2001} T. Otsuka, R. Fujimoto, Y. Utsuno, B. A. Brown, 
  M. Honma, and T. Mizusaki, Phys. Rev. Lett. \textbf{87} (2001), 082502.

\bibitem{otsuka_tensor_2005} T. Otsuka, T. Suzuki, R. Fujimoto, H. Grawe, 
  and Y. Akaishi, Phys. Rev. Lett. \textbf{95} (2005), 232502.

\bibitem{otsuka_tensor_2010} T. Otsuka, T. Suzuki, M. Honma, Y. Utsuno, 
  N. Tsunoda, K. Tsukiyama, and M. H.-Jensen, 
  Phys. Rev. Lett. \textbf{104}, (2010), 012501.

\bibitem{otsuka_threebody_2010} T. Otsuka, T. Suzuki, J. D. Holt, A. Schwenk, 
  and Y. Akaishi, Phys. Rev. Lett. \textbf{105}, (2010) 032501.




%------------------------------------------------------------------------------
% References in Section 2 Shimizu
%------------------------------------------------------------------------------

\bibitem{ppnp_brown} B. A. Brown, Prog. Part. Nucl. Phys. {\bf 47} (2001), 517.

\bibitem{okinawa_shimizu} N. Shimizu, Y. Utsuno, T. Mizusaki,
  T. Otsuka, T. Abe and M. Honma,
  AIP Conf. Proc. {\bf 1355} (2011), 138.

% Primary references for JISP16
\bibitem{Shirokov07} A.~M.~Shirokov, J.~P.~Vary, A.~I.~Mazur and T.~A.~Weber,
                   Phys. Letts. B {\bf 644} (2007), 33; A.~M.~Shirokov, J.~P.~Vary, A.~I.~Mazur, S.~A. Zaytsev 
                   and T.~A.~Weber, Phys. Lett. B {\bf 621} (2005), 96; subroutines to generate this interaction
                   in the relative-center-of-mass HO basis are available at nuclear.physics.iastate.edu


\bibitem{lawson} D.H. Gloeckner and R.D. Lawson, Phys. Lett. 
  \textbf{53B} (1974), 313.

\bibitem{fda} M. Honma, B. A. Brown, T. Mizusaki, and T. Otsuka, 
  Nucl. Phys. A {\bf 704}, (2002), 134c, M. Honma, T. Otsuka, B.A. Brown and 
  T. Mizusaki,  Phys. Rev. C \textbf{65} (2002), 061301(R).

\bibitem{ppnp_schmid} K. W. Schmid, Prog. Part. Nucl. Phys. 
  {\bf 52} (2004), 565.

\bibitem{hybrid_h2} G. Puddu, Eur. Phys. J. A {\bf 34} (2007), 413.

\bibitem{puddu_eqs} G. Puddu, J. Phys. G: Nucl. Part. Phys. {\bf 32} (2006), 321.

\bibitem{num_recipe} Numerical Recipes in Fortran 77, the Art of Scientific Computing,
  2nd ed., Cambridge University Press: Cambridge, (1992).

\bibitem{cg_egido} J. L. Egido, J. Lessing, V. Martin, L. M. Robledo, 
  Nucl. Phys. A \textbf{594} (1995), 70.

\bibitem{mshell64} T. Mizusaki, N. Shimizu, Y. Utsuno, and M. Honma,
  code MSHELL64, unpublished.

\bibitem{horoi_conv} M. Horoi, A. Volya, and V. Zelevinsky,
  Phys. Rev. Lett. {\bf 82} (1999), 2064; M. Horoi, B. A. Brown, and V. Zelevinsky,
  Phys. Rev. C \textbf{67} (2003), 034303.

\bibitem{extrap_2ndorder} T. Mizusaki and M. Imada,
  Phys. Rev. C {\bf 65} (2002), 064319; {\it ibid.}
  {\bf 67} (2003), 041301.

\bibitem{papenbrock_factorize} T. Papenbrock and D. J. Dean,
  Phys. Rev. C \textbf{67} (2003), 051303(R).

\bibitem{shen_zhao} J.J. Shen, Y. M. Zhao, A. Arima,
  and N. Yoshinaga, Phys. Rev. C \textbf{83} (2011), 044322.

\bibitem{imada_pirg} M. Imada and T. Kashima, J. Phys. Soc. Jpn. {\bf 69} (2000), 2723.

\bibitem{fpd6} W.A. Richter, M.G. van der Merwe, R.E. Julies and B.A. Brown, 
  Nucl. Phys. {\bf A523}, 325,  (1991).

\bibitem{shimizu_reordering} N. Shimizu, Y. Utsuno, T. Mizusaki, 
  M. Honma, Y. Tsunoda, and T. Otsuka, 
  Phys. Rev. C,  \textbf{85} (2012), 054301.

\bibitem{ni_coex_mcsm} T. Mizusaki, T. Otsuka, Y. Utsuno, M. Honma
  and T. Sebe, Phys. Rev. C \textbf{59} (1999), R1846.

\bibitem{jun45} M. Honma, T. Otsuka, T. Mizusaki, and M. Hjorth-Jensen,
  Phys. Rev. {\bf C 80} (2009), 064323.

\bibitem{mcsm_tuning} Y. Utsuno, N. Shimizu, T. Otsuka and T. Abe, 
  arXiv:1202.2957 [nucl-th] (2012).
  % submitted to Comp. Phys. Comm.

\bibitem{pfg9b3} M. Honma {\it et al.}, unpublished.

\bibitem{intel_mkl} Intel Math Kernel Library, 
  http://software.intel.com/en-us/articles/intel-mkl/

\bibitem{t2k-todai} T2K Open Supercomputers, 
  http://www.cc.u-tokyo.ac.jp/system/ha8000/


%------------------------------------------------------------------------------
% References in Section 3
%------------------------------------------------------------------------------
\bibitem{Abe:2012wp}
  T.~Abe, P.~Maris, T.~Otsuka, N.~Shimizu, Y.~Utsuno and J.~P.~Vary,
  %``Benchmarks of the ab initio FCI, MCSM and NCFC methods,''
  arXiv:1204.1755 [nucl-th].
  %%CITATION = ARXIV:1204.1755;%%

%Primary references for chiral interactions:
\bibitem{Epelbaum} E. ~Epelbaum, W. ~Gl\"ockle, and Ulf-G. ~Meissner,
Nucl. Phys. A {\bf 637} (1998), 107; {\bf 671} (2000), 295.
\bibitem{N3LO} D. ~R. ~Entem and R. ~Machleidt,
% \bibinfo{journal}{nucl-th/0304018} \textbf{\bibinfo{volume}{}},
Phys. Rev. C {\bf 68} (2003), 041001(R).
% Primary reference to Argonne and Illinois potentials
\bibitem{Wiringa:1994wb}
  R.~B.~Wiringa, V.~G.~J.~Stoks and R.~Schiavilla,
  %``An Accurate nucleon-nucleon potential with charge independence breaking,''
  Phys.\ Rev.\  C {\bf 51} (1995), 38.
  %%CITATION = PHRVA,C51,38;%%
%  Realistic models of pion-exchange three-nucleon interactions
\bibitem{Pieper_3NF}
S.~C.~Pieper, V.~R.~Pandharipande, R.~B.~Wiringa, and J.~Carlson,
Phys. Rev. C {\bf 64} (2001), 014001.
%\Main reference for Illinois Extension to the Fujita-Miyazawa Three-Nucleon Force
\bibitem{Illinois}
S.~C.~Pieper, AIP Conf. Proc. {\bf 1011} (2008), 143.


%Primary references for GFMC - provided by Steven Pieper 
\bibitem{GFMC}
% "Quantum Monte Carlo calculations of excited states in A=6-8 nuclei"
S.~C.~Pieper, R.~B.~Wiringa, and J.~Carlson,
Phys. Rev. C {\bf 70} (2004), 054325; 
% "Quantum Monte Carlo Calculations of Neutron-? Scattering"
K.~M.~Nollett, S.~C.~Pieper, R.~B.~Wiringa, J.~Carlson, and G.~M.~Hale
Phys. Rev. Lett. {\bf 99} (2007), 022502;
S.~C.~Pieper
Proceedings of the International School of Physics "Enrico Fermi", 
Course CLXIX, edited by A. Covello, F. Iachello and R. A. Ricci 
(Societ Italiana di Fisica, Bologna, 2008)  111.
arXiv:0711.1500v1 [nucl-th]; Reprinted in La Rivista del Nuovo Cimento, {\bf 31} (2008), 709;
and references therein.

%Primary reference for CC
\bibitem{CC}
G.~Hagen, T.~Papenbrock and M.~Hjorth-Jensen,
Phys. Rev. Lett. {\bf 104} (2010), 182501
and references therein.

% Okubo-Lee-Suzuki-Okamoto
\bibitem{Lee_Suzuki_Okamoto}
S.~Okubo, Prog. Theor. Phys. {\bf 12} (1954), 603; 
S.~Y.~Lee and K.~Suzuki, Phys. Lett. B, {\bf 91} (1980), 173; 
K.~Suzuki and S.~Y.~Lee, Prog. Theor. Phys. {\bf 64} (1980), 2091; 
K.~Suzuki, R.~Okamoto, Prog. Theor. Phys. {\bf 70} (1983), 439.

\bibitem{RG}
  S.~K.~Bogner, R.~J.~Furnstahl and A.~Schwenk,
  %``From low-momentum interactions to nuclear structure,''
  Prog.\ Part.\ Nucl.\ Phys.\  {\bf 65} (2010), 94.
  %%CITATION = PPNPD,65,94;%%

\bibitem{Maris:2011as}
  P.~Maris, J.~P.~Vary, P.~Navratil, W.~E.~Ormand, H.~Nam and D.~J.~Dean,
  %``Origin of the anomalous long lifetime of 14C,''
  Phys.\ Rev.\ Lett.\  {\bf 106} (2011), 202502.
  %%CITATION = PRLTA,106,202502;%%

\bibitem{Roth}
  R.~Roth,
  %``Importance Truncation for Large-Scale Configuration Interaction
  %Approaches,''
  Phys.\ Rev.\  C {\bf 79} (2009), 064324; 
  %%CITATION = PHRVA,C79,064324;%%
%\cite{Roth:2011vt}
%\bibitem{Roth:2011vt}
  R.~Roth, S.~Binder, K.~Vobig, A.~Calci, J.~Langhammer and P.~Navratil,
  %``Ab Initio Calculations of Medium-Mass Nuclei with Normal-Ordered Chiral
  %NN+3N Interactions,''
  arXiv:1112.0287 [nucl-th].
  %%CITATION = ARXIV:1112.0287;%%

\bibitem{Dytrych} 
T.~Dytrych, K.~D.~Sviratcheva, C.~Bahri, J.~P.~Draayer and J.~P.~Vary, 
% Evidence for Symplectic Symmetry in Ab Initio No-Core Shell Model Results for Light Nuclei,
Phys. Rev. Lett. {\bf 98} (2007), 162503;
T.~Dytrych, K.~D.~Sviratcheva,  C.~Bahri, J.~P.~Draayer and J.~P.~Vary, 
% Highly deformed modes in the ab initio symplectic no-core shell model,
J. Phys. G. {\bf 35} (2008), 095101;
T.~Dytrych, K.~D.~Sviratcheva, J.~P.~Draayer, C.~Bahri and J.~P.~Vary, 
% Ab initio symplectic no-core shell model,
J. Phys. G. {\bf 35} (2008), 123101.

% Benchmarks in the no-core MCSM
\bibitem{okinawa_tabe} T. Abe, P. Maris, T. Otsuka, N. Shimizu, Y. Utsuno,
  and J. P. Vary,   AIP Conf. Proc. {\bf 1355} (2011), 173.

\bibitem{Puddu}
  G.~Puddu,
  %``An efficient method to evaluate energy variances for extrapolation
  %methods,''
  arXiv:1201.0600 [nucl-th].
  %%CITATION = ARXIV:1201.0600;%%


% No-core MCSM calculation for 10Be and 12Be
\bibitem{Liu2011}
L.~Liu, T.~Otsuka, N.~Shimizu, Y.~Utsuno and R.~Roth, 
Phys. Rev. C in press.
% arXiv:1105.2983 [nucl-th].


%\cite{Utsuno:2012vm}
\bibitem{Utsuno:2012vm}
  Y.~Utsuno, N.~Shimizu, T.~Otsuka and T.~Abe,
  %``Efficient computation of Hamiltonian matrix elements between non-orthogonal
  %Slater determinants,''
  arXiv:1202.2957 [nucl-th].
  %%CITATION = ARXIV:1202.2957;%%


% Primary references for the NCFC 
\bibitem{NCFC} 
P.~Maris, J.~P.~Vary, A.~M.~Shirokov, Phys. Rev. C {\bf 79} (2009), 014308;
P.~Maris, A.~M.~Shirokov and J.~P.~Vary, 
% "Ab initio nuclear structure simulations: the speculative 14F nucleus",
Phys. Rev. C {\bf 81} (2010), 021301(R);
C.~Cockrell, J.~P.~Vary and P.~Maris, 
%  "Lithium isotopes within the ab initio no-core full configuration approach"
arXiv:1201.0724.



%%%%%%%%%%%%%%%%%%%%%%%%%%%%%%%%%%%%%%%%%%%%%%%%%%%%%%%%%%%%%%%%%%%%%%%%%%%%%%%
% References in Section 3-2 Yoshida
\bibitem{yoshida:2000wp}
  R. B. Wiringa, S. C. Pieper, J. Carlson, and V. R. Pandharipande, 
  Phys. Rev. C {\bf 62} (2000), 014001.
\bibitem{yoshida:2005im}
  N. Itagaki, H. Masui, M. Ito, and S. Aoyama, Phys. Rev. C {\bf 71} (2005), 064307.
\bibitem{yoshida:2012NCFC}
  C. Cockrell, J. P. Vary and P. Maris, arXiv:1201.0724 [nucl-th].

% \bibitem{MCSM} YOSHIDA TRANSFORMED COEFFICIENTS ``MCSM''??

%%%%%%%%%%%%%%%%%%%%%%%%%%%%%%%%%%%%%%%%%%%%%%%%%%%%%%%%%%%%%%%%%%%%%%%%%%%%%%%

%------------------------------------------------------------------------------
% References in Section 4
%------------------------------------------------------------------------------

\bibitem{Nowacki}
 S.~M.~Lenzi, F.~Nowacki, A.~Poves, and K.~Sieja, 
% {\em Island of inversion around $^{64}$Cr\/},
 \PRC{82,2010,054301}.

\bibitem{A3DA}
 M. Honma {\it et al.}, unpublished.

\bibitem{GXPF1A}
 M.~Honma, T.~Otsuka, B.~A.~Brown, and T.~Mizusaki, 
% {\em Shell-model description of neutron-rich $pf$-shell nuclei with 
%      a new effective interaction GXPF1\/},
 Eur.\ Phys.\ J.\ A \textbf{25} (2005), s01, 499.

\bibitem{Gmatrix}
 M.~Hjorth-Jensen, T.~T.~S.~Kuo, and E.~Osnes, 
% {\em Realistic effective interactions for nuclear systems\/},
 \PR{261,1995,125}.

\bibitem{Hjorth-Jensen}
 M.~Hjorth-Jensen, private communication.

\bibitem{nudat}
 National Nuclear Data Center, information extracted from the NuDat 2 database, http://www.nndc.bnl.gov/nudat2/ 

\bibitem{BE2}
 B.~Pritychenko, J.~Choquette, M.~Horoi, B.~Karamy, and B.~Singh, 
% {\em An Update of $B(E2)$ Evaluation for $0^+_1\rightarrow 2^+_1$ Transitions in Even-Even Nuclei near $N\sim Z\sim 28$\/},
 arXiv:1102.3365v2.

% COMMENT NOT REFFERED BY Y.TSUNODA?
\bibitem{78Ni}
% M.-G.~Porquet, and O.~Sorlin, arXiv:1201.1097v1.  
  M.-G.~Porquet, and O.~Sorlin, \PRC{85,2012,014307}

\bibitem{utsuno_espe}
  Y. Utsuno, T. Otsuka, T. Mizusaki, and M. Honma, 
  Phys. Rev. C \textbf{60} (1999), 054315.



%%%%% appendix

\bibitem{kcomputer} K computer, http://www.aics.riken.jp/en/

\bibitem{ringschuck} P. Ring and P. Schuck, 
  {\it The Nuclear Many-Body Problem}, Springer-Verlag, New York, 1980.

\bibitem{puddu_eff_variance}  G. Puddu, arXiv:1201.0600 [nucl-th].


\end{thebibliography}
\end{document}